\documentclass[11pt, twoside, a4paper]{article}
\usepackage{feynmp}
\usepackage{amsmath, amsthm, amssymb, array, dsfont, mathtools}
\usepackage[margin=1.1in]{geometry}
\usepackage[pdftex]{graphicx}
\usepackage{bm, comment, tikz, tikz-cd, setspace, mathrsfs, microtype, siunitx, booktabs, cancel, caption, 
colortbl, csquotes, grffile, multirow, listings, pgfplots, xcolor}
\usepackage{subfig}
\usepackage{authblk}
\usepackage{cite}
\usepackage{framed}
\usepackage{ulem}

\allowdisplaybreaks
\pgfplotsset{compat=1.12}

\definecolor{MyDarkBlue}{rgb}{0.15,0.15,0.45}
\usepackage[linktocpage=true]{hyperref}
\hypersetup{
colorlinks=false,
citecolor=MyDarkBlue, 
linkcolor=MyDarkBlue,
urlcolor=MyDarkBlue,
}

\makeatletter
\renewcommand*{\@textcolor}[3]{%
 \protect\leavevmode
 \begingroup
 \color#1{#2}#3%
 \endgroup
}
\makeatother

\newcommand{\del}{\partial}

\newcommand{\fr}[1]{\mathfrak{#1}}
\newcommand{\mc}[1]{\mathcal{#1}}

\newcommand{\wt}{\widetilde}
\newcommand{\wh}{\widehat}

\newcommand{\red}[1]{\textcolor{red}{#1}}

\renewcommand{\a}{\alpha}
\renewcommand{\b}{\beta}
\newcommand{\g}{\gamma}
\renewcommand{\d}{\delta}
\newcommand{\e}{\epsilon}

\newcommand{\h}{\eta}
\renewcommand{\th}{\theta}

\newcommand{\m}{\mu}
\newcommand{\n}{\nu}
\newcommand{\x}{\xi}

\renewcommand{\r}{\rho}

\newcommand{\f}{\phi}

\renewcommand{\o}{\omega}
\newcommand{\G}{\Gamma}
\newcommand{\D}{\Delta}

\renewcommand{\S}{\Sigma}
\newcommand{\F}{\Phi}

\renewcommand{\O}{\Omega}

\newcommand{\ol}{\overline}

\newcommand{\La}{\mathcal{L}}

\newcommand{\be}{\begin{equation}}
\newcommand{\ee}{\end{equation}}

\numberwithin{equation}{section}

\begin{document}
\unitlength = 1mm
\setlength{\parskip}{1em}

\title{\fontsize{15}{18}\selectfont\textbf{Ladder Symmetries of Black Holes and de Sitter Space: \\Love Numbers and Quasinormal Modes}}
\author[1]{Roman Berens}
\author[1]{Lam Hui}
\author[2]{Zimo Sun}
\affil[1]{\fontsize{9}{12}\selectfont \textit{Center for Theoretical Physics, Department of Physics, Columbia University, New York, NY 10027}}
\affil[2]{\fontsize{9}{12}\selectfont \textit{Princeton Gravity Initiative, Princeton University, Princeton, NJ 08544, USA}}

\date{}
\maketitle

\begin{abstract}
In this note, we present a synopsis of geometric symmetries for (spin 0) perturbations around (4D) black holes and de Sitter space. For black holes, we focus on static perturbations, for which the (exact) geometric symmetries have the group structure of SO(1,3). The generators consist of three spatial rotations, and three conformal Killing vectors obeying a special {\it melodic} condition. The static perturbation solutions form a unitary (principal series) representation of the group. The recently uncovered ladder symmetries follow from this representation structure; they explain the well-known vanishing of the black hole Love numbers. For dynamical perturbations around de Sitter space, the geometric symmetries are less surprising, following from the SO(1,4) isometry. As is known, the quasinormal solutions form a non-unitary representation of the isometry group. We provide explicit expressions for the ladder operators associated with this representation. In both cases, the ladder structures help connect the boundary condition at the horizon with that at infinity (black hole) or origin (de Sitter space), and they manifest as contiguous relations of the hypergeometric solutions.
\end{abstract}

\newpage
\tableofcontents

\newpage

\section{Introduction}
\label{intro}

Geometric symmetries---isometries and conformal isometries---have always played an important role in field theories. In this note, we wish to apply the geometric viewpoint to two well known phenomena: the vanishing of the black hole Love numbers, and the existence of quasinormal modes in de Sitter space. The reason for focusing on them is that the underlying symmetries are exact, and they share a number of common features as we shall see.

It was recently pointed out that static black hole perturbations enjoy certain exact symmetries, termed ladder symmetries \cite{Hui:2021vcv} (see also \cite{BenAchour:2022uqo, Katagiri:2022vyz}).
They help explain the well known vanishing of the Love numbers, which characterize black holes' response to static tidal fields \cite{1972ApJ...175..243P, Martel:2005ir, Fang:2005qq, Damour:2009va, Damour:2009vw, Binnington:2009bb, Kol:2011vg, Landry:2015cva, Landry:2015zfa, Gurlebeck:2015xpa, Porto:2016pyg, Poisson:2020mdi, LeTiec:2020spy, LeTiec:2020bos,Chia:2020yla,Goldberger:2020fot, Hui:2020xxx, Charalambous:2021mea}.\footnote{In \cite{Charalambous:2021kcz} a different set of symmetries, in a near-zone approximation, were used to explain the vanishing of the Love numbers. The representation-theoretic approach adopted in this paper is similar to \cite{Charalambous:2021kcz}, but the symmetries are different. In particular, ours do not involve time. It is possible to extend our exact ladder symmetries to perturbations of small non-zero frequencies (for which the symmetries are only approximate), but it requires using a different near-zone approximation from that of \cite{Charalambous:2021kcz}; the relevant near-zone approximation is the one proposed by \cite{1973JETP...37...28S}. For further discussion, see \cite{Hui:2022vbh}.}
For simplicity, in this note, we focus on spin 0 perturbations. It can be shown that the spin 1 and 2 perturbations are obtainable from the spin 0 ones by simple (derivative) operations;\footnote{The spin raising (and lowering) operations are possible for uncharged black holes. See \cite{Hui:2021vcv}.}
the relevant spin 0 perturbations $\Phi$ obey:
\begin{eqnarray}
\label{4Dbox}
\Box_{\rm 4D} \Phi = 0 \quad ; \quad ds_{\rm 4D}^2 = -\frac{\D}{r^2} \, dt^2 + \frac{r^2}{\D} \, dr^2 + r^2 (d\theta^2 + {\,\rm sin}^2 \theta \,d\varphi^2) ,
\end{eqnarray}
where $\Box_{\rm 4D}$ is defined with respect to the 4D metric given. Here, we consider a general spherically symmetric (Reissner-Nordst\"om) black hole, with $\D \equiv r^2 - r_S r + r_Q^2$, where $r_S \equiv 2 GM$ and $r_Q^2 \equiv G Q^2$, with $M$ being the black hole mass and $Q$ being the charge. For static $\Phi$, it can be shown that the equation of motion can be equivalently expressed as:
\begin{eqnarray}
\label{3Dbox}
\Box \Phi = 0 \quad ; \quad ds^2 = dr^2 + \D (d\theta^2 + {\,\rm sin}^2 \theta \,d\varphi^2) ,
\end{eqnarray}
where $\Box$ is now defined with respect to the 3D effective metric given. It is worth stressing this 3D metric is not the same as the spatial part of the 4D starting point. The 3D effective metric has a vanishing Cotton tensor (but non-vanishing and non-constant Ricci), and has 10 conformal Killing vectors (CKVs).\footnote{That the 3D effective metric is conformally flat is not surprising---this is true for $\Delta$ being any function of $r$. What is less trivial is the condition on some of the CKVs \eqref{melodicDef0}.}
Three of these are Killing vectors (KVs), generating the expected rotational symmetries. The rest at first sight do not seem useful, since the scalar $\Phi$ is not conformally coupled. Surprisingly, three of them turn out to generate symmetries for $\Phi$. In \cite{Hui:2021vcv}, this was attributed to the fact that the 3D effective metric is conformally related to an Euclidean AdS space. In this note, we wish to point out a different way of viewing the surprise: it has to do with the fact that each of these three conformal Killing vectors $X^i$ obey a special condition which we term {\it melodic}:
\begin{eqnarray}
\label{melodicDef0} 
\Box \nabla_i X^i = 0 .
\end{eqnarray}
where the derivatives are defined with respect to the 3D effective metric. A main objective of this note is to explain the relevance of this condition for generating symmetries. As we shall see, the three rotational KVs and the three melodic CKVs form an $\fr{so}(1,3)$ algebra. The solutions to Eq. (\ref{3Dbox}), organized by spherical harmonics, form a representation of the algebra. The representation structure is what gives rise to the ladder symmetries pointed out by \cite{Hui:2021vcv}.

This representation understanding of static black hole perturbations can be usefully compared with the representation understanding of dynamical de Sitter perturbations. It was first noticed in \cite{Anninos:2011af} that the de Sitter quasinormal modes solved by \cite{Lopez-Ortega:2006aal} can be organized into representations of $\text{SL}(2,\mathbb R)$. Later in \cite{Ng:2012xp, Jafferis:2013qia}, these quasinormal modes were found to form a representation of the isometry algebra $\fr{so}(1,4)$, which was generalized to higher dimensions and higher spins by  \cite{Sun:2020sgn}.
 We shall review that story, and provide expressions for the myriad of ladder operators, analogous to those given for black holes.

A common theme in both stories is the connection of boundary conditions. In both cases, the relevant radial equation (after separation of variables) is a second order ordinary differential equation. Close to any boundary, there are in general two asymptotic behaviors. For instance, for the black hole, the static $\Phi$ approaches either a constant or a logarithm (or combinations thereof) close to the horizon; it approaches either $r^\ell$ or $1/r^{\ell+1}$ (or combinations thereof) at infinity, where $\ell$ is the angular momentum quantum number. The Love number surprise, so to speak, has to do with the fact that the regular solution, where $\Phi$ approaches a constant at the horizon, has only the ``growing'' behavior $r^\ell$ at infinity, i.e., the ``decaying'' behavior $1/r^{\ell+1}$ is completely absent. Likewise, for dynamical perturbations in the static patch of de Sitter space, $\Phi$ approaches either an ingoing or an outgoing wave (or combinations thereof) at the horizon; it approaches either $r^\ell$ or $1/r^{\ell+1}$ (or combinations thereof) at the origin. Quasinormal modes are those solutions at special frequencies, such that
$\F$ is purely outgoing at the horizon, and purely $r^\ell$ at the origin. 

As we shall see, the ladder structures help explain this single-asymptote behavior, i.e., a single asymptotic behavior at one boundary is connected with another single asymptotic behavior at a different boundary. At the level of the radial solutions for $\Phi$, which are hypergeometric functions, this phenomenon is associated with what are called connection formulas. The ladder structures manifest as a generalized form of contiguous relations governing hypergeometric functions, which we shall spell out.

It is worth stressing that much of this paper is a review of known results. Our goal is thus a modest one: to view both the black hole Love numbers and the de Sitter quasinormal modes through the same geometric/ representation-theoretic lens. For black holes, the surprise, if there is one, is the relevance of melodic CKVs. For de Sitter space, there is less of a surprise since the symmetries arise straightforwardly from KVs. In both cases, organizing the solutions in a representation of the symmetry algebra is a helpful way of thinking about the ladder structures. For de Sitter space, we provide expressions for operators (some of them new) for climbing up and down the myriad ladders. In both cases, the ladder operators provide a simple way to understand the single-asymptote behavior of the solutions.

Quasinormal modes exist also for black holes of course \cite{Chandrasekhar:1975zza, Kokkotas:1999bd, Berti:2004md, Berti:2009kk}. 
The reason we do not focus on them is
that it is not known what exact symmetries govern them, if any, beyond the usual rotational symmetries. There have been interesting recent developments pointing out approximate symmetries \cite{Raffaelli:2021gzh, Hadar:2022xag, Kapec:2022dvc, Kehagias:2022ndy}.
We shall discuss them in Section \ref{discussion}.

The paper is organized as follows.
In Section \ref{BH}, we highlight the importance of the melodic CKVs in giving the static scalar perturbations around a spherically symmetric black hole the symmetry algebra of $\fr{so}(1,3)$. The perturbation solutions form a unitary representation, from which the ladder structures and symmetries of \cite{Hui:2021vcv} follow. The arguments for the vanishing of the Love numbers are reviewed here.
In Section \ref{dS}, we apply the same geometric approach to a dynamical scalar of an arbitrary mass (or coupling to Ricci) in de Sitter space. Multiple ladders exist, due to the larger symmetry algebra. Explicit expressions for the ladder operators are provided, and they are used to deduce the quasinormal spectrum, and the single-asymptote nature of the quasinormal modes. 
If the scalar is conformally coupled, more symmetries exist, the consequences of which are derived from a representation perspective.
In Section \ref{discussion}, we summarize our results and discuss future directions.
Appendices \ref{app:CKV_S3} and \ref{app:CKV_dS} give the (C)KVs of the relevant spaces and their commutation relations. \ref{app:solutions} contains the exact solutions for the equations of motion. \ref{app:hyper_id} shows the algebraic origin of the ladders from various contiguous relations for hypergeometric functions. 
\ref{app:commutator} shows the mutual consistency among three different interpretations of the commutator.
\ref{app:MCKV} explores properties of the melodic CKVs. \ref{app:ladder_identities} collects various identities obeyed by the ladder operators. \ref{app:horizontal} puts conditions on the existence of certain ``horizontal'' symmetries useful for arguments in the main text.
\ref{app:identities} presents the general form of some useful identities for first order differential operators of a certain type. Finally, \ref{app:ladder_extensions} shows how to combine various ladder operators to move flexibly throughout the space of solutions.

\section{Black Hole Ladder}
\label{BH}

\subsection{Geometric Symmetries} 
\label{BHgeom}

Our starting point is the 3D effective metric \eqref{3Dbox}:
\begin{eqnarray}
\label{eq:ds3D}
ds^2 = dr^2 + \D (d\theta^2 + {\,\rm sin}^2 \theta \,d\varphi^2) ,
\end{eqnarray}
in which the static scalar $\Phi$ lives.
It originates from thinking about a massless, static scalar in the background of a general, spherically symmetric, i.e., Reissner--Nordstr\"{o}m, black hole \eqref{4Dbox}.\footnote{The inner/outer horizons are at $r = r_\pm \equiv (r_S \pm \sqrt{r_S^2 - 4 r_Q^2})/2$. 
Note $\D \equiv r^2 - r_S r + r_Q^2$ can be rewritten as $(r - r_+) (r - r_-)$, and we have $r_+ + r_- = r_S$ and $r_+r_- = r_Q^2$.}
This 3D space has a vanishing Cotton tensor, and thus is conformally flat. It has $10$ (C)KVs, forming an $\fr{so}(1,4)$ algebra.\footnote{A full account of the $10$ (C)KVs is given in Appendix \ref{app:CKV_S3}.} Three of these are the familiar KVs for rotation:
\begin{align}
J_1^i \del_i &= -\sin\varphi \, \del_\th 
- \cos\varphi\cot\th \, \del_\varphi\nonumber \\
J_2^i \del_i &= \cos\varphi \, \del_\th 
- \cot\th \sin\varphi \, \del_\varphi\nonumber \\
J_3^i \del_i &= \del_\varphi.
\label{eq:BH_KV}
\end{align}
The remaining seven---CKVs---do not at first sight seem useful, since $\Phi$ is not conformally coupled. 
Surprisingly, this na\"{i}ve expectation is false. It turns out three CKVs are special:
\begin{align}
K_1^i \del_i &= \D(r) \sin\th \cos\varphi \, \del_r 
+ \frac{1}{2}\D'(r)
\left(-\cos\th\cos\varphi \, \del_\th
+ \csc\th\sin\varphi \, \del_\varphi\right)\nonumber \\
K_2^i \del_i &= \D(r) \sin\th \sin\varphi \, \del_r
- \frac{1}{2}\D'(r)
\left(\cos\th\sin\varphi \, \del_\th
+ \csc\th\cos\varphi \, \del_\varphi\right)\nonumber \\
K_3^i \del_i &= \D(r) \cos\th \, \del_r 
+ \frac{1}{2}\D'(r)\sin\th \, \del_\th ,
\label{eq:BH_MCKV}
\end{align}
where $\Delta' \equiv \partial_r \Delta = 2r - r_S$. They obey what we call the {\it melodic} condition:
a CKV $X$ is melodic if it satisfies
\be
\label{melodicDef}
\Box \nabla_i X^i = 0 .
\ee
For the case at hand, $\nabla_i$ and $\Box$ are defined with respect to the 3D space \eqref{eq:ds3D}. 
It can be shown that a melodic CKV generates a symmetry for a massless scalar, even if the scalar
is not conformally coupled. The symmetry transformation it effects is:
\be \label{eq:deltaXckv} 
\d_X \F = X^i \nabla_i \F 
+ \frac{d-2}{2d} \nabla_i X^i \F,
\ee
where $d = 3$ in the case at hand. 
A proof is provided in the following inset.
\begin{framed}
{\small
\vspace{-.15cm}
\noindent Consider the following action for
a scalar $\Phi$ in $d$ dimensions:
\be S = \frac{1}{2}\int d^d x \sqrt{|g|} \, \left(\F \Box \F - \x R \F^2\right),\ee
where the coupling $\xi$ need {\it not} take the conformal value
$\x_c \equiv \frac{d-2}{4(d-1)}$. (We keep $\xi$ as a free parameter for the sake of generality;
for our black hole application, we are interested in $\xi=0$.)
It can be shown that under the transformation
\be \d_X \F = X^i \nabla_i \F + \frac{d-2}{2d} \nabla_i X^i \F, \label{eq:delta_X}\ee
where $X$ is a CKV, the action transforms as
\begin{align}
\d_X S &= \int d^d x \sqrt{|g|} \left(
\frac{1}{2} \nabla_i \left(X^i \F \Box \F - \x X^i R\F^2\right)
- \frac{d-1}{d}\left(\x-\x_c\right)
\left(\Box\nabla_i X^i \right)\F^2\right)\\
&= -\frac{d-1}{d}\left(\x-\x_c\right) 
\int d^d x \sqrt{|g|} \, 
\left(\Box\nabla_i X^i\right) \F^2 , 
\label{eq:deltaS_CKV}
\end{align}
where we have dropped the boundary term in the last line. Thus, the CKV $X$ defines a symmetry of the action if $\xi = \xi_c$ (conformal coupling), or if $\Box \nabla_i X^i = 0$ ($X$ is a melodic CKV). A KV can
be thought of a trivial or degenerate case, since $\nabla_i X^i = 0$. 

The above discussion is at the level of the action. At the level of the equation of motion, it can be shown that if $X$ is a CKV, then
\be (\Box - \xi R) \d_X \F 
= \ol{\d}_X (\Box - \xi R) \F 
- \frac{2(d-1)}{d}\left(\x-\x_c\right)
\Box \nabla_i X^i,
\label{Main_eq:laplacian_sym}
\ee
where
\be \ol{\d}_X \F = X^i \nabla_i \F 
+ \frac{d+2}{2d} \nabla_i X^i \F.
\label{Main_eq:delta_X_bar}
\ee
}
In other words, if $\xi=\xi_c$ (conformal coupling), or $\Box \nabla_i X^i = 0$ ($X$ is a melodic CKV), $\delta_X$ passes through the equation of motion operator $\Box - \xi R$ (and becomes $\ol{\d}_X$).
This is how the symmetry manifests itself at the level of the equation of motion.
Incidentally, it can further be shown that for the melodic CKVs $X_1, X_2, \ldots$ (or for $\xi=\xi_c$), 
\be (\Box - \xi R) \d_{X_1} \cdots \d_{X_n} \Phi
= \ol{\d}_{X_1} \cdots \ol{\d}_{X_n} (\Box - \xi R) \Phi .
\ee
These are related to higher spin symmetries \cite{Eastwood:2002su}, which can arise from conformal Killing tensors (CKTs), and in this case the CKTs are decomposable, i.e., they can be expressed as a symmetric product of CKVs. Further properties of melodic CKVs are found in Appendix \ref{app:MCKV}.
\end{framed}

The upshot is that, somewhat surprisingly, a static, massless scalar $\Phi$ around a spherically symmetric 
black hole enjoys a large amount of {\it exact} symmetries: generated by three KVs (the $J_i$) and three CKVs (the $K_i$). 
They obey an $\fr{so}(1,3)$ algebra:
\begin{align}
\left[J_i, J_j\right] = -\e_{ijk} J_k, 
\qquad
\left[\bar K_i, \bar K_j\right] 
= \e_{ijk} J_k,
\qquad
\left[J_i, \bar K_j\right] 
= -\e_{ijk} \bar K_k,
\end{align}
where $\bar K_i$ is simply a rescaled version of $K_i$: $\bar K_i \equiv 2K_i/\sqrt{r_S^2 - 4 r_Q^2}$.\footnote{
In the flat space limit ($r_S, r_Q \rightarrow 0$), it is better not to rescale. $K_i$ generates a special conformation transformation, and $[K_i, K_j] = 0$. The extremal limit is similar, with $r$ replaced by $r - r_S/2$. In fact, the exact solutions are simply $z^{\ell}$ and $z^{-(\ell+1)}$, with $z = r$ in the flat space case and $z = r - r_S/2$ in the extremal case. In both cases the algebra of the CKVs instead becomes the Lie algebra of the Euclidean group $\text{ISO}(3)$.} 
Here, we are abusing the notation somewhat: the brackets $\left[ \cdot , \cdot \right]$ can be interpreted as the Lie bracket of the enclosed vectors or as the commutator of the symmetry transformations effected by the enclosed vectors. (The symmetry transformation is not equal to the Lie derivative in the case of the $K_i$, because
$\nabla_i K^i \ne 0$. See \eqref{eq:deltaXckv}.)
The interesting point is that the algebra is exactly the same regardless of the interpretation. For a proof, see Appendix \ref{app:commutator}.

The symmetry structure is identical to that of the Lorentz algebra, with the $K_i$ playing the role of boosts. To see in detail how $J_i$ and $K_i$ act on the solutions of $\Box\F = 0$, it is useful to decompose $\Phi$ in spherical harmonics, with the solutions in harmonic space labeled by $\ell, m$: 
$\F_{\ell m} (r,\th,\varphi) \equiv \f_\ell(r) Y_{\ell m}(\th,\varphi)$, where the radial function $\f_\ell$ satisfies
\be \del_r \left(\D\del_r \f_\ell\right) - \ell(\ell+1)\f_\ell = 0.
\label{eq:RN_EoM}
\ee
We have suppressed the $m$ label on the radial function because the equation of motion does not depend on it, due to spherical symmetry of the background.

Imagine stacking the solutions $\phi_\ell Y_{\ell m}$, labeled by $\ell$ and $m$, into a giant (infinite) column vector. We can form a representation of the $\fr{so}(1,3)$ algebra by asking how the generators act on this column vector of solutions. It is clear how rotations act: for instance, under the action of $J_1$ and $J_2$, spherical harmonics of different $m$ mix while keeping $\ell$ the same. This gives the familiar ladder structure of the $J_i$, where they are represented by anti-hermitian matrices.\footnote{In this work we use the convention that a unitary representation of the Lie group means the Lie algebra generators are represented as anti-hermitian matrices.} From the point of view of the $\fr{so}(3)$ of rotations, the representation is reducible to blocks of size $(2\ell+1) \times (2\ell+1)$, but the $K_i$ will mix up solutions of different $\ell$ (and also $m$ for $K_1$ and $K_2$).

Let us illustrate with the action of $K_3$. Its effect on $\Phi$, according to \eqref{eq:deltaXckv}, is:
\begin{align}
\d_{K_3} \Phi = \left[ \D(r) \cos\th \, \del_r 
+ \frac{1}{2}\D'(r) \sin\th \, \del_\th 
+ \frac{1}{2}\D'(r) \cos\th \right] \Phi .
\end{align}
Expanding in spherical harmonics, this implies (without assuming $\phi_\ell$ solves \eqref{eq:RN_EoM}):
\be 
\d_{K_3}\left(\f_\ell Y_{\ell m}\right)
= -f(\ell) D_\ell^+ \f_\ell Y_{\ell+1,m}
+ f(\ell-1) D_\ell^- \f_\ell Y_{\ell-1,m} ,
\label{eq:RN_delta_K3}
\ee
where\footnote{Useful identities are: 
\begin{align}
\cos\th \, Y_{\ell m} &= f(\ell) Y_{\ell+1,m} 
+ f(\ell-1) Y_{\ell-1,m}\nonumber \\
\sin\th \, \del_\th Y_{\ell m}
&= \ell f(\ell) Y_{\ell+1,m} 
- (\ell+1) f(\ell-1) Y_{\ell-1,m},
\end{align} }
\be 
f(\ell) = \sqrt{\frac{(\ell-m+1)(\ell+m+1)}{(2\ell+1)(2\ell+3)}} .
\label{eq:f}\ee
Here, the operators $D^\pm_\ell$ are defined by:
\begin{align}
\label{Ddef}
D_\ell^+ \equiv -\D \del_r - \frac{\ell+1}{2}\D', \qquad
D_\ell^- \equiv \D \del_r - \frac{\ell}{2} \D' .
\end{align}
Their significance can be deduced as follows. The transformation generated by $K_3$, being a symmetry, maps solutions to solutions. Thus, if $\phi_\ell Y_{\ell m}$ is a solution, $\delta_{K_3} (\phi_\ell Y_{\ell m}) $ must be also. Looking at the right hand side of \eqref{eq:RN_delta_K3}, the first term multiplies $Y_{\ell+1, m}$; thus $D^+_\ell \phi_\ell$ must be the radial solution at the $\ell+1$ level, assuming that all $\phi_\ell$ are regular at $r_+$\footnote{The eq. (\ref{eq:RN_EoM}) has two linearly independent solutions, one regular at $r=r_+$, and the other one irregular at $r=r_+$. The operators $D_\ell^\pm$ clearly preserve the (ir)regularity condition at $r=r_+$. For the discussion of representation structure here, we always consider regular solutions. See the next subsection for an extensive discussion on (ir)regularity of the solutions to $\Box\Phi=0$.}.
 Similarly, the second term multiplies $Y_{\ell-1, m}$; thus $D^-_\ell \phi_\ell$ must be the radial solution at the $\ell-1$ level. In other words, $D^\pm_\ell$ acting on the  radial solution $\phi_\ell$ at level $\ell$ maps it to $\phi_{\ell+1}$ or $\phi_{\ell-1}$, i.e., they are raising and lowering operators.\footnote{One can abstract out from this reasoning a vertical ladder symmetry, following the terminology of \cite{Hui:2021vcv}: express the scalar action as sum in harmonic space, and focus on two levels $\ell$ and $\ell-1$; it can be shown $\delta \phi_{\ell} = D^+_{\ell-1} \phi_{\ell-1}$ and $\delta \phi_{\ell-1} = - D^-_\ell \phi_\ell$ is a symmetry. This pairing of levels is reminiscent of supersymmetry. See discussion in \cite{Hui:2021vcv}.}

So far we have shown that the space of solutions to $\Box\Phi=0$, which is spanned by all $\phi_\ell Y_{\ell m}$, furnishes an infinite dimensional irreducible representation of $\rm{SO} (1,3)$. We can demonstrate the unitarity of this representation for the regular solutions using the following inner product:
\begin{align}\label{inner}
    (\Phi_1, \Phi_2) = \int_{S^2}\, d\Omega\,  \Phi^*_1(r_+,\theta, \varphi) \Phi_2(r_+,\theta, \varphi)~,
\end{align}
which is effectively the standard $L^2$ inner product on $S^2$. Choosing the convenient normalization:\footnote{This normalization is inappropriate in both the flat space and extremal limits, as explained in footnote 6. However, the existence of the ladder operators and the subsequent arguments about the vanishing of the Love numbers remain valid.
}
\begin{align}
\label{D+D-}
D^+_\ell \phi_\ell = N_{\ell+1} \phi_{\ell+1}, 
\qquad D^-_\ell \phi_\ell = 
N_\ell \phi_{\ell-1} ,
\end{align}
where 
\be N_\ell \equiv \frac{\ell}{2} \sqrt{r_S^2 - 4 r_Q^2}=\frac{\ell}{2}(r_+-r_-),\ee
we find the regular solutions are orthogonal and obey
\begin{align}
    (\phi_\ell Y_{\ell m}, \phi_{\ell} Y_{\ell m})=  \left|\phi_\ell(r_+)\right|^2 =\left|\phi_0\right|^2.
\end{align}
Note that regularity at the horizon forces $\f_0$ to be a constant. The rotation generators are anti-hermitian automatically, and it can be easily checked that all the $K_i$ are also anti-hermitian, for example 
\begin{align}
    (\Phi_1, K_3\Phi_2)=\frac{r_+- r_-}{2}\int_{S^2}\, d\Omega\,  \Phi^*_1(r_+,\theta, \varphi)\partial_\theta\left[ \sin\theta\Phi_2(r_+,\theta, \varphi)\right]=- (K_3\Phi_1, \Phi_2)~.
\end{align}
Altogether, the regular solutions of $\Box\Phi=0$ equipped with the inner product \eqref{inner} carry a unitary irreducible representation of $\rm SO(1, 3)$.

To identify this representation, we can use the SO$(1,3)$ Casimir. We have
\begin{align}
\d^{ij} \d_{J_i} \d_{J_j}
&= \nabla^2_{S^2}\\
\d^{ij} \d_{K_i} \d_{K_j}
&= \D \del_r \left(\D \del_r\right) + \frac{\left(\D'\right)^2}{4}\nabla^2_{S^2}
- \frac{r_S^2- 4r_Q^2}{4},
\end{align}
where $\nabla^2_{S^2}$ is the angular Laplacian.\footnote{The operator $\Box$ according to the 3D effective metric is
$\Box = \D^{-2}\left(\d^{ij} \d_{K_i} \d_{K_j} 
- \frac{r_S^2 - 4r_Q^2}{4}\d^{ij} \d_{J_i} \d_{J_j} 
+ \frac{r_S^2 - 4r_Q^2}{4}\right)$. The 4D operator $\Box_{\rm 4D}$, restricted to static perturbations, is $(\Delta/r^2)\Box$.} 
Acting on solutions in harmonic space, they give:
\begin{align} 
\d^{ij} \d_{J_i} \d_{J_j}\left(\f_\ell Y_{\ell m}\right)
&= -\ell(\ell+1) \f_\ell Y_{\ell m}\nonumber \\
\d^{ij} \d_{K_i} \d_{K_j}\left(\f_\ell Y_{\ell m}\right)
&= -\left(\ell(\ell+1) + 1\right)\frac{r_S^2 - 4r_Q^2}{4} \f_\ell Y_{\ell m}.
\end{align}
Thus, the Casimir $\d^{ij} \d_{J_i} \d_{J_j} - \d^{ij} \d_{\bar K_i} \d_{\bar K_j} = 1$. 
The set of solutions therefore form the  principal series representation $\mathcal{P}_{1, 0}$ \cite{Sun:2021thf}.\footnote{We thank Alessandro Podo for discussions on this point.}

Before we close this subsection, let us introduce a generalization of 
\eqref{eq:RN_delta_K3}, following the treatment by \cite{Sun:2020sgn}. Schematically, \eqref{eq:RN_delta_K3} takes the form $\delta_{K_3} (\Phi_{\ell m}) \sim \Phi_{\ell+1 \, m} + \Phi_{\ell-1 \, m}$, where $\Phi_{\ell m}$ represents $\phi_\ell Y_{\ell m}$ and $\F_{\ell\pm 1,m}$ represents $D^\pm_\ell \f_{\ell m} Y_{\ell\pm1,m}$. Starting from $\ell=0, m=0$, we see that $\delta_{K_3} (\Phi_{00}) \sim \Phi_{10}$ (the term $\Phi_{-1 0} = 0$ due to the form of $f$ in \eqref{eq:RN_delta_K3}). Doing this again yields $\delta_{K_3} \delta_{K_3} (\Phi_{00}) \sim \Phi_{20} + \Phi_{00}$. Moving $\Phi_{00}$ to the left, we see that there is a modification of $\delta_{K_3} \delta_{K_3}$ that would raise the $\ell=0$, $m=0$ solution to the $\ell=2$, $m=0$ solution. This is in fact general. There exists a tensor $c^{\ell m}$ such that
\be
\label{buildPhiellm}
\sum_{i_1, \ldots, i_\ell} c^{\ell m}_{i_1 \cdots i_\ell} \delta_{K_{i_1}} \cdots \delta_{K_{i_\ell}} \Phi_{00} \propto \Phi_{\ell m},
\ee
This procedure maps the $\Phi_{00}$ solution to a solution at an arbitrary $\ell$ and $m$, $\Phi_{\ell m} = \phi_\ell Y_{\ell m}$, up to some normalization constant (the constant depends on $\ell$ and $m$ if one adopts the convention in \eqref{D+D-}). The $\ell$-index tensor $c^{\ell m}$ is completely symmetric and traceless, and turns out to be exactly the same as that used to construct spherical harmonics:
\be 
\label{cYellm}
r^\ell Y_{\ell m} (\theta, \varphi) =  \sum_{i_1, \ldots, i_\ell} c^{\ell m}_{i_1 \cdots i_\ell} x^{i_1} \cdots x^{i_\ell},
\ee
where the $x^i$ represent the Cartesian coordinates. A proof is given in Appendix \ref{app:identities}. Equation \eqref{buildPhiellm} gives an economical way to build the entire representation starting from the $\fr{so}(3)$ singlet, a spherically symmetric solution. 

\subsection{Ladder Structure and Love Numbers}

To understand what the geometric symmetries have to say about the Love numbers (one for each $\ell$), we need to work in harmonic space, and
learn what the ladder structure tells us about the nature of the solutions \cite{Hui:2021vcv}.

The equation of motion for the radial function $\phi_\ell$ \eqref{eq:RN_EoM}
can be rewritten as 
\begin{align}
\label{Hphi}
H_\ell \phi_\ell = 0, \qquad H_\ell \equiv -\D\left(\del_r\left(\D\del_r\right) - \ell(\ell+1)\right) .
\end{align}
Using $D^+_\ell$ and $D^-_\ell$ defined in \eqref{Ddef}, it can be shown that
\be H_{\ell+1} D_\ell^+ = D_\ell^+ H_\ell, \qquad
H_{\ell-1} D_\ell^- = D_\ell^- H_\ell.
\label{eq:ladder}\ee
Thus if $\f_\ell$ is a solution to the equation of motion for level $\ell$, $D^\pm_\ell\f_\ell$ is a 
solution at level $\ell \pm 1$.
This reconfirms that $D^\pm_\ell$ indeed act as raising and lowering operators.
Furthermore, these operators satisfy the relations:
\begin{align}
\label{eq:DDcommute}
& D_{\ell+1}^- D_\ell^+ - D_{\ell-1}^+ D_\ell^- 
= \frac{2\ell+1}{4}\left(r_S^2 - 4r_Q^2\right)\\
\label{eq:HDD}
& H_\ell = D_{\ell+1}^- D_\ell^+ - \frac{(\ell+1)^2}{4}
\left(r_S^2 - 4r_Q^2\right) = D_{\ell-1}^+ D_\ell^- - \frac{\ell^2}{4}
\left(r_S^2 - 4r_Q^2\right).
\end{align}
The last expression implies that if $H_\ell \f_\ell = 0$, both $D^-_{\ell+1} D^+_\ell \f_\ell$ and $D^+_{\ell-1} D^-_\ell \f_\ell$ are proportional to $\f_\ell$. 
Note that the constant $r_S^2 - 4r_Q^2 = (r_+ - r_-)^2$.

As explained above, the ladder structure originates from geometric symmetries.
There is another way to see that such a structure must be present: by recalling
contiguous relations of hypergeometric functions. 
Observe that $\phi_\ell$ obeys a second order ordinary differential equation with three regular singular points, at $r = r_-$, $r_+$, and $\infty$. Its solutions are hypergeometric functions $_2 F_1 (a, b, c)$ whose arguments are related to $\ell$. Contiguous relations relate hypergeometric functions with neighboring arguments, i.e., they take the form of some operator acting on $_2 F_1 (a, b, c)$, giving another $_2 F_1$ with $a, b$, and/or $c$ incremented by $1$, and by extension incrementing $\ell$ by $1$ also. Details can be found in \cite{Hui:2021vcv} and summarized in Appendix \ref{app:hyper_id}. From this point of view, it is not surprising that the ladder operators $D^\pm_\ell$ exist.

How is the ladder structure helpful for the Love number problem? Let us recall what the
problem is. Consider what the equation of motion \eqref{eq:RN_EoM} tells us
about the asymptotic behavior of $\phi_\ell$ at infinity and at the outer horizon.
As $r \rightarrow \infty$, it is straightforward to see that $\phi_\ell$ goes as $r^\ell$ or $1/r^{\ell+1}$. As $r \rightarrow r_+$, $\phi_\ell$ goes as a constant or as $\ln[(r - r_+)/r_+]$. The phenomenon of vanishing Love numbers has to do with the fact that the solution that is regular at the horizon (i.e., going as a constant as $r \rightarrow r_+$) has a purely $r^\ell$ asymptotic at large $r$. The na\"{i}ve expectation, that it has a mixture of ``growing'' $r^\ell$ and ``decaying'' $1/r^{\ell+1}$ at large $r$, does not hold. This is the surprise of the vanishing Love number, defined as the ratio of the decaying tail to the growing one. It is a problem of connecting asymptotic behaviors at two different boundaries: how come regularity at the horizon is connected with purely ``growing'' behavior at infinity?

Let us look at \eqref{eq:RN_EoM} and 
\eqref{eq:HDD}. They tell us 
\be
H_\ell \phi_\ell = \left[ D^+_{\ell-1} D^-_{\ell} - 
\frac{\ell^2}{4} (r_S^2 - 4 r_Q^2) \right] \phi_\ell = 0
\ee
The constant term in $H_\ell$ vanishes for $\ell = 0$, i.e., $H_0 \phi_0 = D^+_{-1} D^-_0 \phi_0 = 0$. It suggests one possible solution is given by a $\phi_0$ that is annihilated by $D^-_0$:
\be D^-_0 \phi_0 = 0 .\ee
This is helpful, because we have turned a second order problem (i.e., $H_0$ involves two derivatives) into a first order one (i.e., $D^-_0$ involves only one derivative). Thus, imposing regularity at the horizon is no longer expected to yield generically an admixture of growing and decaying tail at infinity. Indeed, $D^-_0$ is simply $\Delta \partial_r$, and
so $D^-_0 \phi_0 = 0$ implies $\phi_0$ is a constant: this is clearly regular at the horizon and purely ``growing'' at infinity (i.e., no $1/r$ tail). Moreover, starting from any solution $\phi_0$, it is simple to construct a solution at level $\ell$ by climbing the ladder:
\be \label{phi0phiell} \f_\ell \propto D^+_{\ell-1} \cdots D^+_1 D^+_0 \phi_0 .\ee
Plugging in the $\phi_0 =$ constant solution, and recalling the form of $D^+_\ell = (r^2 - r_S r + r_Q^2) \partial_r - (\ell+ 1) (r - r_S/2)$, it is 
straightforward to see that the resulting $\phi_\ell$ approaches a constant at the horizon and is purely growing, i.e., it has the $r^\ell$ asymptotic behavior with no $1/r^{\ell+1}$ tail (indeed, it cannot have any negative power tail).

Strictly speaking, the above argument for the vanishing of the Love numbers requires an extra comment. All we have shown is that there is a solution that is regular at the horizon and purely growing at infinity. How do we know there is not an independent solution that is also regular at the horizon but contains a decaying tail at infinity? We know that cannot happen, because an independent solution cannot also approach a constant at the horizon as its leading behavior, i.e., it must approach the other possible asymptotic behavior $\ln[(r-r_+)/r_+]$ and is irregular at the horizon.\footnote{On the other hand, it is possible to have two independent solutions which both approach the logarithm as their leading behavior at the horizon, 
because one of them
could have the constant asymptote hidden in its subleading terms. But it is impossible to have two independent solutions both approaching the constant as their leading behavior.} \footnote{
\label{footnote_phi0}
Indeed, one can deduce further statements about such irregular solutions. Going back to $H_0 \phi_0 = D^+_{-1} D^-_0 \phi_0 = 0$: instead of looking for a solution that satisfies $D^-_0 \phi_0 = 0$ (i.e., the constant solution), we could look for a different solution such that $D^-_0 \phi_0 \ne 0$, but $D^+_{-1} (D^-_0 \phi_0) = 0$. Recalling the form of $D^+_{-1}$, we see that this implies $D^-_0 \phi_0$ equals a non-zero constant. Solving this tells us $\phi_0 \propto {\,\rm ln\,} [(r-r_+)/(r-r_-)]$. This solution diverges logarithmically as $r \rightarrow r_+$, and goes as $1/r$ at large $r$. It can be shown that raising it by $D^+_\ell$ operators yield solutions at higher $\ell$ that retain the logarithmic divergence at the horizon. Their large $r$ behavior consists of a mixture of growing and decaying terms.
}

There is another way to connect boundary conditions, by exploiting conservation. The observation is that there is another kind of symmetry at play, one at each level $\ell$, termed a horizontal symmetry by \cite{Hui:2021vcv}, following the work of \cite{Compton:2020cjx}. Note that:
\be 
\label{eq:Q0H0}
[Q_0 , H_0] = 0, \quad \text{where} \quad Q_0 \equiv D_0^- = \Delta \partial_r.
\ee
(This can be quickly understood by examining \eqref{eq:ladder} for $\ell = 0$ and noting that $H_{-1} = H_0$.) It can be shown that $\delta \phi_0 = Q_0 \phi_0$ is a symmetry at the level of the action. By climbing up and down the ladder, it can be further shown that:
\be 
\label{eq:QellHell}
[Q_\ell, H_\ell] = 0, \quad \text{where} \quad Q_\ell \equiv D^+_{\ell-1} Q_{\ell-1} D^-_\ell .
\ee
The transformation $\delta \phi_\ell = Q_\ell \phi_\ell$ is also a symmetry at the level of the action (written as a sum over modes in harmonic space). See Appendix \ref{app:horizontal} for a proof. From the horizontal symmetry at each level $\ell$, a Noether current can be derived, which in our static context takes the form $\partial_r J^r_\ell = 0$, i.e., the $r$ component of the current is a conserved quantity, in the sense of being $r$-independent. Carrying out this procedure is straightforward though a bit cumbersome, and it is in fact simpler to identify the conserved quantity by inspection. Observing that the $\ell=0$ equation of motion is $\partial_r (\Delta \partial_r \phi_0) = 0$, we see that 
\be \label{eq:P0here} P_0 = \D \del_r \f_0 \ee
is conserved on the equation of motion. 
For any higher $\ell$, it is natural to define
\be \label{eq:Pellhere} 
P_\ell \equiv \D \del_r\left(D_1^- D_2^- \cdots D_\ell^- \f_\ell\right),\ee
which is also conserved on the equation of motion, i.e., $\partial_r P_\ell = 0$, since the string of lowering operators merely serve to lower $\phi_\ell$ to $\phi_0$ (up to a constant). It is worth noting that actually carrying out the Noether procedure would yield $J^r_\ell = P_\ell^2$. Obviously, its conservation is equivalent to the conservation of $P_\ell$.\footnote{It can be seen that $P_0$ is in fact the Wronskian between $\phi_0$ and the constant solution. Normally, the conservation of Wronskian follows from a trivial symmetry of a linear system: one can always add to the field a solution of the equation of motion. What is curious here is that a less trivial symmetry, i.e. $\delta \phi_0 = Q_0 \phi_0$, gives rise to a conserved quantity whose square root turns out to be $P_0$.
}
Henceforth, slightly abusing the terminology, we shall refer to $P_\ell$ as a conserved {\it charge}.\footnote{Suppose $\phi_\ell$ in \eqref{eq:Pellhere} is built by raising from some $\phi_0$, i.e., $A_\ell \phi_\ell = D^+_{\ell-1} \cdots D^+_0 \phi_0$, where 
$A_\ell$ follows from a string of normalization factors \eqref{D+D-}:
$A_\ell = \prod_{i = 1}^{\ell} N_i = \ell! \left(r_S^2 - 4r_Q^2\right)^{\ell/2} / 2^\ell$.
It can be shown that
\be P_\ell = \frac{\ell!}{2^\ell} 
\left(r_S^2 - 4 r_Q^2\right)^{\ell/2} P_0.\ee
This follows from noting that $P_\ell = A_\ell^{-1} \Delta \partial_r (D^-_1 \cdots (D^-_\ell D^+_{\ell-1}) D^+_{\ell-2} \cdots D^+_0 \phi_0)$. We highlight by parentheses $D^-_\ell D^+_{\ell-1}$, and observe that to its right is essentially $\phi_{\ell-1}$ which solves $H_{\ell-1} \phi_{\ell-1} = 0$. This tells us $[D^-_\ell D^+_{\ell-1} - \ell^2 (r_S^2 - 4 r_Q^2)/4] \phi_{\ell-1} = 0$, so we can replace $D^-_\ell D^+_{\ell-1}$ by $\ell^2 (r_S^2 - 4 r_Q^2)/4$. Repeating the procedure to lower rungs of the ladder yields the above expression. 
}

How do we use the conservation of $P_\ell$ to tackle the problem of connecting boundary conditions? Following \cite{Hui:2021vcv}:
first of all, observe that if $\phi_\ell$ approaches a constant at the horizon, $D^-_1 \cdots D^-_\ell \phi_\ell$ would produce something regular at the horizon, and so $P_\ell = 0$ when evaluated at the horizon, due to the factor of $\Delta$ in \eqref{eq:Pellhere}.
In other words, the regular solution has a vanishing $P_\ell$ charge. Conversely, if $\phi_\ell$ approaches the divergent logarithm $\ln [(r - r_+)/r_+]$ at the horizon, $P_\ell \ne 0$ when evaluated at the horizon. Thirdly, if $\phi_\ell \sim 1/r^{\ell+1}$ at large $r$, the form of $D^-_\ell$ tells us that $P_\ell \ne 0$ when evaluated at large $r$. Conservation of $P_\ell$ thus tells us a solution with a purely decaying tail $1/r^{\ell+1}$ at large $r$ must diverge at the horizon, consistent with Bekenstein's no-hair theorem \cite{Bekenstein:1995un}. Lastly, let's deduce the large $r$ behavior of the solution that is regular at the horizon, which we shall call $\phi_\ell^{\rm reg.}$. As noted above, $P_\ell = 0$ at the horizon for $\phi_\ell^{\rm reg.}$, and thus by conservation, $P_\ell = 0$ at large $r$ too. Looking at \eqref{eq:Pellhere}, this implies $D^-_1 \cdots D^-_\ell \phi_\ell^{\rm reg.}$ at large $r$, when written as a power series, starts at most at $r^0$. Crucially, the subleading terms cannot have a $1/r$ contribution; otherwise $P_\ell$ would be nonzero.
It is useful at this point to recall $D^-_1 \cdots D^-_\ell \phi_\ell^{\rm reg.}$ is by construction a solution at the $\ell=0$ level, for which we know the most general solution is a superposition of a constant and a logarithm (see footnote \ref{footnote_phi0}). Importantly, the logarithm has a $1/r$ contribution at large $r$. This means the only way for $\phi_\ell^{\rm reg.}$ to yield a zero $P_\ell$ at large $r$ is for $D^-_1 \cdots D^-_\ell \phi_\ell^{\rm reg.}$ to be just a constant. Inverting this, we can say $\phi_\ell^{\rm reg.}$ has to be equal to $D^+_{\ell-1} \cdots D^+_0$ acting on a constant. By inspecting the form of $D^+_\ell$, we conclude $\phi_\ell^{\rm reg.}$ goes as $r^\ell$ at large $r$ and has no negative power law tail, and in particular no $1/r^{\ell+1}$ contribution.

\section{de Sitter Ladders}
\label{dS}

We now turn our attention to de Sitter space. We shall work in the static patch, with the metric given by
\be ds^2 = -\frac{\D}{r^2} \, dt^2 + \frac{r^2}{\D} \, dr^2 + r^2 d\O^2,\ee
where now $\D \equiv r^2\left(1-r^2/L^2\right)$. We consider a scalar field in de Sitter space with an arbitrary coupling $\x$ to the Ricci scalar. The equation of motion is
\be 
\label{eq:eomdS}
\left(\Box - \x R\right)\F = 0.\ee
Since $R = \frac{d(d-1)}{L^2}$ is a constant (with $d = 4$ in our case), one could also interpret $\x R$ as mass squared. The scalar is conformally coupled if $\x = \x_c \equiv \frac{d-2}{4(d-1)}$. It turns out to be more convenient to work with the parameter
\be \a \equiv \frac{(d-1)^2}{4} - \x \,d(d-1) .\ee
Conformal coupling corresponds to $\alpha = \frac{1}{4}$, while masslessness corresponds to $\alpha= \frac{(d-1)^2}{4}$.

Let us emphasize again many of the results below are not new. The quasinormal spectrum was worked out in \cite{Lopez-Ortega:2006aal}, and its relation to the geometric symmetries was pointed out by \cite{Anninos:2011af, Ng:2012xp, Jafferis:2013qia, Sun:2020sgn}, though the expressions for some of the ladder operators, and the conserved charges (for the conformally coupled case), are new. The main motivation for including a discussion of de Sitter space is to emphasize the many similar features shared with the black hole problem. The way the ladder structures emerge out of representations is the same. And the way asymptotic behaviors at separate boundaries can be connected---relevant for Love numbers for black holes, and quasinormal modes for de Sitter space---is also the same.

\subsection{Geometric Symmetries}
\label{dSgeom}

For de Sitter space, there does not appear to be the analog of melodic CKVs, i.e., CKVs that obey condition \eqref{melodicDef} in a non-trivial way. Thus, we focus on the KVs. We shall comment on the extra symmetries that a conformally coupled scalar enjoys in Section \ref{conformallycoupled}. 

The isometry algebra of de Sitter space is $\fr {so}(1,4)$. For the purpose of understanding the ladder structures, it is helpful to see their explicit form in static patch coordinates:
\begin{align}
J_1^\m \del_\m &= -\sin\varphi \, \del_\th 
- \cos\varphi\cot\th \, \del_\varphi \nonumber \\
J_2^\m \del_\m &= \cos\varphi \, \del_\th 
- \cot\th \sin\varphi \, \del_\varphi \nonumber \\
J_3^\m \del_\m &= \del_\varphi \nonumber \\
D^\m \del_\m &= -L \del_t \nonumber \\
P_1^\m \del_\m &= 
e^{-t/L} \left(-\frac{r^2}{\sqrt{\D}}
\sin\th \cos\varphi\,\del_t 
+ L\frac{\sqrt{\D}}{r}
\left(\sin\th \cos\varphi \,\del_r 
+ \frac{1}{r} 
\cos\th \cos\varphi\,\del_\th 
- \frac{1}{r} \csc\th\sin\varphi \,\del_\varphi\right)\right) \nonumber \\
P_2^\m \del_\m &= 
e^{-t/L} \left(-\frac{r^2}{\sqrt{\D}}
\sin\th \sin\varphi\,\del_t 
+ L\frac{\sqrt{\D}}{r}\left(
\sin\th \sin\varphi \,\del_r 
+ \frac{1}{r} \cos\th \sin\varphi \,\del_\th 
+ \frac{1}{r} \csc\th\cos\varphi \,\del_\varphi\right)\right) \nonumber \\
P_3^\m \del_\m &= 
e^{-t/L} \left(-\frac{r^2}{\sqrt{\D}}
\cos\th \,\del_t 
+ L\frac{\sqrt{\D}}{r}
\left(\cos\th \,\del_r 
- \frac{1}{r} \sin\th \,\del_\th \right)
\right) \nonumber \\
K_1^\m \del_\m &= 
e^{t/L} \left(\frac{r^2}{\sqrt{\D}}
\sin\th \cos\varphi\,\del_t 
+ L\frac{\sqrt{\D}}{r}
\left(\sin\th \cos\varphi \,\del_r 
+ \frac{1}{r} 
\cos\th \cos\varphi\,\del_\th 
- \frac{1}{r} 
\csc\th\sin\varphi \,\del_\varphi\right)\right) \nonumber \\
K_2^\m \del_\m &= 
e^{t/L} \left(\frac{r^2}{\sqrt{\D}}
\sin\th \sin\varphi\,\del_t 
+ L\frac{\sqrt{\D}}{r}\left(
\sin\th \sin\varphi \,\del_r 
+ \frac{1}{r} \cos\th \sin\varphi \,\del_\th 
+ \frac{1}{r} \csc\th\cos\varphi \,\del_\varphi\right)\right) \nonumber \\
K_3^\m \del_\m &= 
e^{t/L} \left(\frac{r^2}{\sqrt{\D}}
\cos\th \,\del_t 
+ L\frac{\sqrt{\D}}{r}
\left(\cos\th \,\del_r - \frac{1}{r} \sin\th \,\del_\th \right)
\right) . \label{eq:dS_KV}
\end{align}
Defining:
\be 
\begin{split}
M_{ij} = \e_{ijk} J_k, \quad
M_{0i} = \frac{1}{2} \left(P_i - K_i\right), \quad
M_{i4} = \frac{1}{2} \left(P_i + K_i\right), \quad
M_{04} = D, \label{eq:dS_to_M}
\end{split}
\ee
the commutation relations are summarized by
\be \left[M_{AB}, M_{CD}\right]
= \h_{AD} M_{BC} + \h_{BC} M_{AD} 
- \h_{AC} M_{BD} - \h_{BD} M_{AC},
\label{eq:so_CR}
\ee
with $\h \equiv {\rm diag\,} (-1, 1, 1, 1, 1)$. 

In particular, the $J_i$ commute with $D$ which is simply time translation. Thus, it is natural to label the solutions by $\o, \ell, m$ where $i\o L$ is the eigenvalue of $D$ (i.e., the solution's dependence on $t$, $\th$, and $\varphi$ takes the form $e^{-i\omega t} Y_{\ell m} (\th, \varphi)$). As we shall see, it turns out to be convenient to label the solutions by $\g$ in place of $\o$:
\be
\label{gammaDef}
\g \equiv i \o L - \ell .
\ee
Thus, applying the separation of variables, solutions to \eqref{eq:eomdS} take the form of 
$\Phi_{\g\ell m} (t, r, \th, \varphi) \equiv \phi_{\g\ell} (r) Y_{\ell m} (\th, \varphi) e^{-i\o_{\g\ell} t}$.\footnote{Our notation, in which $\o$ carries $\g$ and $\ell$ labels, might seem unmotivated. It will make more sense in a moment. For now, it is perfectly acceptable to think of the solutions as labeled instead by $\o$, $\ell$, and $m$, i.e., $\Phi_{\o\ell m} = \phi_{\o\ell} Y_{\ell m} e^{-i\o t}$.}
They form a representation of the $\fr{so}(1,4)$ isometry algebra.

As in the case of the black hole, $J_1$ and $J_2$ mix solutions of different $m$ without affecting $\ell$ and $\g$. The $K_i$ and $P_i$ have non-trivial commutators with $J_i$ and $D$, and will mix different $\ell$ and $\g$ (and in general $m$ too).\footnote{The complete set of commutation relations is shown in Appendix \ref{app:CKV_dS}.}
For instance, $K_3$, $P_3$ and $D$ form an $\fr{sl}(2,\mathbb{R})$ subalgebra:
\be \left[D, P_3 \right] = P_3, \quad
\left[D, K_3\right] = -K_3, \quad
\left[K_3, P_3\right] = 2 D . \ee
(The same statement can be made with $K_3, P_3$ replaced by $K_1$, $P_1$ or $K_2$, $P_2$.) Thus $K_3$ and $P_3$ can be thought of as raising and lowering operators, incrementing the eigenvalue of $D$, i.e., $i\o L$. This can also be seen explicitly by noting that $K_3$ and $P_3$ contain factors of $e^{\pm t/L}$. Since $K_3$ and $P_3$ do not commute with $J_1$ and $J_2$, they at the same time mix up solutions with different $\ell$ (without affecting $m$):
\begin{align}
\d_{P_3}\left(\f_{\g\ell} Y_{\ell m} 
e^{-i\o_{\g\ell} t}\right) 
&= 
L e^{-(i\o_{\g\ell} L + 1)t/L}
\left(
f(\ell) E^+_{\g\ell} \f_{\g\ell} Y_{\ell+1,m} 
- f(\ell-1) F^+_{\g\ell} \f_{\g\ell} Y_{\ell-1,m}
\right)
\nonumber 
\\
\d_{K_3}\left(\f_{\g\ell} Y_{\ell m} 
e^{-i\o_{\g\ell} t}\right) 
&= 
L e^{-(i\o_{\g\ell} L - 1)t/L}
\left(
f(\ell) F^-_{\g\ell} \f_{\g\ell} Y_{\ell+1,m} 
- f(\ell-1) E^-_{\g\ell} \f_{\g\ell} Y_{\ell-1,m}
\right),
\label{eq:dS_delta_K3}
\end{align}
where $f(\ell)$ is again given by \eqref{eq:f}, and we have defined the operators
\begin{align} 
\label{EFdef}
E_{\g\ell}^+ &\equiv \frac{\sqrt{\D}}{r} \del_r 
- \ell\frac{\sqrt{\D}}{r^2} 
+ i\o_{\g\ell} \frac{r^2}{L\sqrt{\D}} \nonumber \\
E_{\g\ell}^- &\equiv -\frac{\sqrt{\D}}{r} \del_r 
- (\ell+1)\frac{\sqrt{\D}}{r^2} 
+ i\o_{\g\ell} \frac{r^2}{L\sqrt{\D}}\nonumber \\
F^+_{\g\ell} &\equiv -\frac{\sqrt{\D}}{r} \del_r 
- (\ell+1)\frac{\sqrt{\D}}{r^2} 
- i\o_{\g\ell} \frac{r^2}{L\sqrt{\D}}\nonumber \\
F^-_{\g\ell} &\equiv \frac{\sqrt{\D}}{r} \del_r 
- \ell\frac{\sqrt{\D}}{r^2} 
- i\o_{\g\ell}\frac{r^2}{L\sqrt{\D}}.
\end{align}
By virtue of being symmetries, $K_3$ and $P_3$ map solutions to solutions. Running an argument similar to the black hole case, we see that $E^\pm_{\g\ell}$ effects $\ell \rightarrow \ell \pm 1$ and
$i\o_{\g\ell} L \rightarrow i\o_{\g\ell}L \pm 1$, and
$F^\pm_{\g\ell}$ effects $\ell \rightarrow \ell \mp 1$ and
$i\o_{\g\ell} L \rightarrow i\o_{\g\ell}L \pm 1$. Notice how the action of $E^\pm_{\g\ell}$ can be alternatively described as effecting $\ell \rightarrow \ell \pm 1$ while keeping $i\o_{\g\ell} L - \ell$ fixed.
This is what motivates the introduction of the symbol $\g$ in \eqref{gammaDef}, i.e., $E^\pm_{\g\ell}$ effects $\ell \rightarrow \ell \pm 1$ without changing $\g$.

Is there a way to effect a shift in $i\o_{\g\ell} L$, without changing $\ell$ (or $m$)? The answer is yes, and unsurprisingly, it involves spherically symmetric combinations of the $K_i$ and $P_i$:
\begin{align}
\d^{ij} \d_{P_i}\d_{P_j} \left(\f_{\g\ell} Y_{\ell m} 
e^{-i\o_{\g\ell}t}\right) 
&= -L^2 G^+_{\g\ell}\f_{\g\ell} 
Y_{\ell m} e^{-(i\o_{\g\ell}L+2)t/L}
\nonumber 
\\
\d^{ij} \d_{K_i}\d_{K_j} \left(\f_{\g\ell} Y_{\ell m} 
e^{-i\o_{\g\ell}t}\right) 
&= -L^2 G^-_{\g\ell}\f_{\g\ell} 
Y_{\ell m} e^{-(i\o_{\g\ell}L-2)t/L} ,
\label{eq:G_from_K}
\end{align}
where $G^\pm_{\g\ell}$ is given by:
\begin{align} 
\label{Gdef}
G^+_{\g\ell} \equiv E^+_{\g+2,\ell-1} F^+_{\g\ell}, 
\qquad 
G^-_{\g\ell} \equiv E^-_{\g-2,\ell+1} F^-_{\g\ell}.
\end{align}
In other words, $\delta^{ij} \delta_{P_i} \delta_{P_j}$ 
(or $G^+_{\g\ell})$
effects $i\o_{\g\ell} L \rightarrow i\o_{\g\ell} L + 2$, while $\delta^{ij} \delta_{K_i} \delta_{K_j}$ 
(or $G^-_{\g\ell})$ effects $i\o_{\g\ell} L \rightarrow i\o_{\g\ell} L - 2$.
Equivalently, we can also say $G^\pm_{\g\ell}$ effects $\gamma \rightarrow \gamma \pm 2$ without changing $\ell$. As before, all of these ladders can also be seen as arising from contiguous relations for the hypergeometric solutions, which are shown in Appendix \ref{app:hyper_id}.


With the above ladder operators defined, it is easy to see how one could construct a representation of $\fr{so}(1,4)$. 
Start from a solution labeled by $i\o_{\g\ell} L$ and $\ell$, or equivalently $\gamma$ and $\ell$ (we suppress $m$ since incrementing it is straightforward using $J_1$ and $J_2$); use $E^\pm_{\g\ell}$ to increment $\ell$ by $\pm 1$ without changing $\gamma$, or use $G^\pm_{\g\ell}$ to increment $\gamma$ by $\pm 2$ without changing $\ell$. 
This way, we fill out a whole table of solutions, which we will also refer to as states.
So far, the only restriction is that $\ell \ge 0$, i.e., the table of $(\g, \ell)$ has $\ell$ taking all possible non-negative integer values, and $\g = \g_0 + 2n$ for any integer $n$ and some base value $\g_0$ (with $\g_0$ arbitrary for the moment). A further restriction on $\gamma$ will come later when we discuss quasinormal modes. Incidentally, since $\gamma$ in general has a non-zero real part, and therefore $i\o_{\g\ell} L$ as well, we have a non-unitary representation (e.g. the matrix representation for time-translation $D$ is not anti-hermitian).

We close this subsection with a description of an economical method to build the entire representation of $\fr{so}(1,4)$, given by \cite{Sun:2021thf}. (This mirrors what was given at the end of Section \ref{BHgeom}.) Taking $c^{\ell m}$ to be the traceless symmetric tensor associated with $Y_{\ell m}$ \eqref{cYellm}, we have
\begin{align}
c^{\ell m}_{i_1\cdots i_\ell} 
\d_{P_{i_1}} \cdots \d_{P_{i_\ell}}
\F_{\g00}(t,r,\th,\varphi)
&\propto \F_{\g \ell m}(t,r,\th,\varphi)\nonumber \\
c^{\ell m}_{i_1\cdots i_\ell} 
\d_{K_{i_1}} \cdots \d_{K_{i_\ell}}
\F_{\g00}(t,r,\th,\varphi)
&\propto \F_{\g-2\ell,\ell m}(t,r,\th,\varphi),
\end{align}
up to some irrelevant normalization constant.
Proofs are once again in Appendix \ref{app:identities}. 
In this language, \eqref{eq:G_from_K} can be recast as:
\begin{align}
\d^{ij} \d_{P_i}\d_{P_j} \F_{\g \ell m}(t,r,\th,\varphi)
&\propto \F_{\g+2, \ell m}(t,r,\th,\varphi)\nonumber \\
\d^{ij} \d_{K_i}\d_{K_j} \F_{\g\ell m}(t,r,\th,\varphi)
&\propto \F_{\g-2,\ell m}(t,r,\th,\varphi).
\end{align}
For further operators connecting different solutions, see
Appendix \ref{app:ladder_extensions}.

\subsection{Ladder Structures and Quasinormal Modes}

The upshot of the last subsection is that de Sitter space has multiple ladders: besides a ladder in $\ell$, there is also a ladder in $\gamma$, defined as $i\omega L - \ell$.
(A ladder in $m$ is present as well, but is no different from the familiar one of spherical harmonics.) Our next task is to learn what the ladder structures tell us about the nature of the quasinormal solutions.

After the separation of variables, the radial function $\phi_{\g\ell} (r)$ obeys
\be \wt{H}_{\g\ell} \phi_{\g\ell} = 0, 
\qquad
\wt{H}_{\g\ell} \equiv
- \frac{r^2}{L^2\Delta} \left(\del_r\left(\D \del_r\right)
+ \frac{\o_{\g\ell}^2r^4}{\D} + \left(\alpha - \frac{9}{4}\right)\frac{r^2}{L^2} - \ell(\ell+1) \right).
\ee 
The asymptotic behaviors of $\phi_{\g\ell}$ are as follows: $\phi_{\g\ell}$ goes as $r^\ell$ or $1/r^{\ell+1}$ as $r \rightarrow 0$; $\phi_{\g\ell}$ goes as $\left(1-\frac{r^2}{L^2}\right)^{-i\o_{\g\ell} L/2}$ (outgoing at horizon) or $\left(1-\frac{r^2}{L^2}\right)^{i\o_{\g\ell} L/2}$ (incoming at horizon) as $r \rightarrow L$.
Linear combinations thereof are of course allowed. What is special about quasinormal modes is that they have such special frequencies that only a single asymptotic behavior is realized at each boundary. In particular, the behavior at the origin is purely regular $r^\ell$ and the behavior at the horizon is purely outgoing $\left(1-\frac{r^2}{L^2}\right)^{-i\o_{\g\ell} L/2}$. 
This is reminiscent of the Love number problem in the case of the black hole: regularity at the horizon and purely growing behavior at infinity. Our task in this subsection is to understand this single-asymptote behavior for the quasinormal solutions.

Our strategy is similar to that in the black hole case \eqref{eq:HDD}: write the equation of motion operator $\wt{H}_{\g\ell}$ as a product of raising and lowering operators plus a constant. We shall use $G^\pm_{\g\ell}$ which raises or lowers $\g$ by $2$ while keeping $\ell$ constant. 
For other ways to rewrite the equation of motion, using $E^\pm_{\g\ell}$ or $F^\pm_{\g\ell}$, see Appendix \ref{app:ladder_identities}.

The operators $G^\pm_{\g\ell}$ defined in \eqref{Gdef} are a bit cumbersome to use, because they involve second derivatives. They can be improved by subtracting off $\wt{H}_{\g\ell}$ (or something proportional to it, keeping in mind that 
the objective is to act on solutions, which are annihilated by $\wt{H}_{\g\ell}$): 
\begin{align} 
\mc{G}^+_{\g\ell} &\equiv 
\left(\g+\ell+1\right)^{-1}
\left(G^+_{\g\ell} - \frac{L^2 \Delta}{r^4} \wt{H}_{\g\ell}\right) \nonumber \\
&= -\frac{2r}{L^2}\del_r 
+ \frac{1}{L^2}\left(
-\frac{2(\g+\ell)}{\D}r^2 
+ \frac{(\g-1)(\g+2\ell) + \a-\frac{9}{4}}{(\g+\ell+1)}
\right)
\label{eq:curlyGplus}
\\
\mc{G}^-_{\g\ell} &\equiv 
\left(\g+\ell-1\right)^{-1}
\left(G^-_{\g\ell} - \frac{L^2 \Delta}{r^4} \wt{H}_{\g\ell}\right)\nonumber \\
&= \frac{2r}{L^2}\del_r 
+ \frac{1}{L^2}\left(
-\frac{2(\g+\ell)}{\D}r^2
+ \frac{\g(\g+2\ell+1) + \a - \frac{9}{4}}{(\g+\ell-1)}
\right).
\label{eq:curlyGminus}
\end{align}
It can be shown that
\be 
\wt{H}_{\g+2,\ell} \,\mc{G}^+_{\g\ell} 
= \mc{G}^+_{\g\ell} \,\wt{H}_{\g\ell}, 
\qquad
\wt{H}_{\g-2,\ell} \,\mc{G}^-_{\g\ell} 
= \mc{G}^-_{\g\ell} \,\wt{H}_{\g\ell}.
\label{eq:G_ladder_CR}
\ee
This reconfirms what we already know from geometric arguments, that given a solution $\phi_{\g\ell}$, $\mc{G}^{\pm}_{\g\ell} \phi_{\g\ell}$ is a solution at level $\g\pm 2$ and the same $\ell$.\footnote{
It is also useful to note that
\be
\mc{G}^-_{\g+2,\ell} \mc{G}^+_{\g\ell} 
- \mc{G}^+_{\g-2,\ell} \mc{G}^-_{\g\ell}
= \frac{4}{L^4} \frac{(\g+\ell) 
\left(A_{\g\ell} + B_{\g\ell}\left(\a-\frac{9}{4}\right)
-\left(\a-\frac{9}{4}\right)^2\right)}{(\g+\ell+1)^2(\g+\ell-1)^2}, \nonumber \\
\ee
\begin{align}
A_{\g\ell} &\equiv 
\left(\g(\g + 2\ell) - \ell + 1\right)
\left(\g(\g + 2\ell) + (2 \ell+3)(\ell-1)\right) , \nonumber \\
B_{\g\ell} &\equiv 2(\ell-1) (\ell+2) .
\end{align}
}
The raising and lowering operators above are the same as those in Section 3.2 of \cite{Anninos:2011af}. 

With this setup, we are ready to rewrite $\wt{H}_{\g\ell}$ in terms of $\mc{G}^{\pm}_{\g\ell}$:
\be
\wt{H}_{\g\ell} 
= \frac{L^2}{4}
\left(
\mc{G}^+_{\g-2,\ell} \mc{G}^-_{\g\ell}
- \frac{1}{L^4}\frac{a_{\g\ell} b_{\g\ell}}
{\left(\g+\ell-1\right)^2}\right)
= \frac{L^2}{4} 
\left(\mc{G}^-_{\g+2,\ell} \mc{G}^+_{\g\ell} 
- \frac{1}{L^4}
\frac{a_{\g+2,\ell} b_{\g+2,\ell}}
{\left(\g+\ell+1\right)^2}
\right) ,
\label{eq:Gid}
\ee
where 
\begin{align}
a_{\g\ell} \equiv \left(\g - \g_+ + 2\ell+1\right)
\left(\g - \g_- + 2\ell+1\right),
\qquad
b_{\g\ell} \equiv \left(\g - \g_+\right)\left(\g - \g_-\right),
\label{eq:const_ab}
\end{align}
with
\be \g_\pm \equiv \frac{3}{2} \pm \sqrt{\a}.
\label{eq:gamma_pm}\ee 
Our strategy is similar to the one adopted for the black hole: observe that the equation of motion 
$\wt{H}_{\g \ell} \phi_{\g\ell} = 0$
simplifies if $\g = \g_\pm$ or $\g = \g_\pm - 2\ell - 1$, such that the constant term in $\wt{H}_{\g\ell}$ 
(the one $\propto a_{\g\ell} b_{\g\ell}$) vanishes. 
With these choices for $\g$,
the equation of motion becomes
\be
\label{G+G-}
\mc{G}^+_{\g - 2, \ell} \,\mc{G}^-_{\g\ell} \, \phi_{\g\ell} = 0 ,
\ee
for which a possible solution is 
\be
\label{Gphi}
\mc{G}^-_{\g\ell} \,\phi_{\g\ell} = 0 .
\ee
As before, this reduction of a second order differential equation to a first order one is what lies behind the single-asymptote behavior. Analysis of \eqref{Gphi}, for the choice $\g = \g_\pm$,
shows that the solution
$\phi_{\g_\pm ,\ell}$ approaches $r^\ell$ at the origin, and $\left(1-\frac{r^2}{L^2}\right)^{-(\g_\pm+\ell)/2}$ at the horizon. These match precisely the desired boundary conditions for a quasinormal mode, i.e. regularity at the origin and outgoing at the horizon.\footnote{
We do not need it, but the exact quasinormal solution is
\be 
\f_{\g_\pm,\ell}(r) = \left(\frac{r}{L}\right)^{\ell} 
\left(1-\frac{r^2}{L^2}\right)^{-(\g_\pm+\ell)/2}.
\label{eq:dS_ground_state}
\ee
}
Recalling \eqref{gammaDef}, we see that the quasinormal frequency is given by
\be
i\o_{\g_\pm \ell} L = \g_\pm + \ell .
\ee
Thus, we have identified two quasinormal frequencies (for a given $\ell$), one for $\g_+$ and one for $\g_-$. Each corresponding quasinormal mode serves as some sort of ``ground state''. The idea is to build the ``excited states'' by successively applying $\mc{G}^+$. This is reminiscent of how the simple harmonic oscillator problem is solved algebraically. 

Before doing so, let us remark upon the other possibility $\g = \g_\pm - 2\ell-1$. In that case, it can be shown that \eqref{Gphi} implies the irregular asymptotic behavior $r^{-{(\ell+1)}}$ at the origin, which does not match the desired quasinormal boundary conditions.\footnote{We do not need this one either, but the exact irregular solution is 
\be 
\f_{\g_\pm - 2\ell - 1,\ell}(r) 
= \left(\frac{L}{r}\right)^{\ell+1} 
\left(1-\frac{r^2}{L^2}\right)^{-(\g_\pm - \ell - 1)/2} .
\ee
}
\footnote{To round out the discussion, one could go back to \eqref{G+G-}
$\mc{G}^+_{\gamma - 2,\ell} \mc{G}^-_{\gamma \ell} \phi_{\g\ell} = 0$, and ask what happens if instead of \eqref{Gphi}, one has
$\mc{G}^-_{\g\ell} \phi_{\g\ell} \neq 0$. In that case, by studying what
$\mc{G}^+_{\gamma-2, \ell}$ annihilates,
it can be shown that the choice $\gamma = \gamma_\pm$ yields $\phi_{\g\ell}$ solutions irregular at the origin, whereas the choice
$\g = \g_\pm - 2\ell - 1$ yields solutions that are incoming at the horizon. Neither
satisfies the desired quasinormal boundary conditions.} 

Returning thus to the choice $\gamma = \gamma_\pm$, having found the ``ground state'' $\phi_{\g_\pm , \ell}$
(for given $\ell$), we can raise to a higher $\gamma$ by successively applying the raising operator (recalling that each time, $\gamma$ increments by $2$):
\be \f_{\g_\pm + 2n,\ell} 
\propto
\mc{G}^+_{\g_\pm + 2n - 2,\ell} 
\mc{G}^+_{\g_\pm + 2n-4,\ell} 
\cdots \mc{G}^+_{\g_\pm+2,\ell} \mc{G}^+_{\g_\pm,\ell} \f_{\g_\pm,\ell},
\ee
where $n$ an integer. It can be shown by inspecting the form of $\mc{G}^\pm_{\g\ell}$ that the resulting $\f_{\g_\pm + 2n,\ell}$ has the correct quasinormal boundary conditions, i.e. $\f_{\g_\pm + 2n,\ell}$ approaches $r^\ell$ at the origin, and approaches $\left(1 - \frac{r^2}{L^2}\right)^{-(\gamma_\pm + 2n + \ell)/2}$ at the horizon.
This way, we can construct all the quasinormal modes, covering the whole spectrum:
\be i \o_{\gamma \ell} L = \gamma + \ell = \g_{\pm} + 2n + \ell, 
\label{eq:QNM_freq}\ee
where $n$ is a non-negative integer. One can further show that the $E^+_{\g\ell}$ operators 
(which raises $\ell$) also preserve the quasinormal boundary conditions. So we can even combine $\mc{G}^+_{\g\ell}$ and $E^+_{\g\ell}$ to construct the entire spectrum from the quasinormal solution $\f_{\g_\pm,0}$, which can be regarded as {\it the} ground state.\footnote{There are some exceptional cases where the procedure---constructing the quasinormal modes by raising $\g$ starting from $\phi_{\g_\pm , \ell}$---requires a modification. This subtlety happens only for the $\g_-$ branch: it turns out if $\g_- = -\ell$ (or equivalently, $2\sqrt{\a} = 2\ell+3$), $\mc{G}^+_{\g_-, \ell}$ is proportional to $\mc{G}^-_{\g_-, \ell}$ which means the former cannot be used to raise $\g$ from $\phi_{\g_- , \ell}$ which is annihilated by the latter. To construct $\phi_{\g_- + 2, \ell}$, one could alternatively use $E^+$ instead. For instance, if $\a = 9/4$ (massless scalar) such that $\g_- = 0$, the $\phi_{00}$ solution is annihilated by both $\mc{G}^+_{00}$ and $\mc{G}^-_{00}$. One could instead use $\phi_{20}$ as the ground state from which to build out the quasinormal spectrum at $\ell=0$, by acting successively with $\mc{G}^+$. To find $\phi_{20}$, use the fact that $E^-_{20} \phi_{20} = 0$.
In other words, forget about the state $\phi_{00}$ (its corresponding frequency is zero anyway); instead, start from $\phi_{20}$ which can be obtained by solving $E^-_{20} \phi_{20} = 0$.}
Or more precisely, there is a ground state for $\g_+$, and a ground state for $\g_-$. From each, a whole table of states labeled by $\g = \g_\pm + 2n$ and $\ell$ can be constructed. In other words, they form two separate (non-unitary) representations of $\fr{so}(1,4)$.

\subsection{Conformally Coupled Case}
\label{conformallycoupled}

A conformally coupled scalar ($\xi = \xi_c$
or $\a = 1/4$ in \eqref{eq:eomdS}) enjoys additional symmetries effected by the CKVs.
The five CKVs of de Sitter space are
\begin{align}
\wt{K}_1^\m \del_\m &= 
\frac{\D}{r^2} \sin\th \cos\varphi \, \del_r 
+ \frac{1}{r} \cos\th \cos\varphi \, \del_\th 
- \frac{1}{r} \csc\th \sin\varphi \, \del_\varphi
\nonumber\\
\wt{K}_2^\m \del_\m &= 
 \frac{\D}{r^2} \sin\th \sin\varphi \, \del_r 
+ \frac{1}{r} \cos\th \sin\varphi \, \del_\th 
+ \frac{1}{r} \csc\th \cos\varphi \, \del_\varphi
\nonumber\\
\wt{K}_3^\m \del_\m &= 
\frac{\D}{r^2} \cos\th \, \del_r 
- \frac{1}{r} \sin\th \, \del_\th \nonumber \\
\wt{K}_+^\m \del_\m &= e^{-t/L} \left(\frac{r}{\sqrt{\D}}\del_t - \frac{\sqrt{\D}}{L} \del_r\right) \nonumber\\
\wt{K}_-^\m \del_\m &= e^{t/L} \left(\frac{r}{\sqrt{\D}}\del_t + \frac{\sqrt{\D}}{L} \del_r\right). \label{eq:dS_CKV}
\end{align}
The commutation relations among them and the KVs are given in Appendix \ref{app:CKV_dS}. Together, they form an $\fr{so}(2,4)$ algebra. From it, one can deduce how they act on solutions, labeled by $\g, \ell, m$ as in the previous subsection. We can also roughly guess their effects based on their explicit form given in \eqref{eq:dS_CKV}. For instance, $\wt{K}_3$ does not involve $t$ or $\varphi$, but involves 
$r$ and $\theta$. Thus we expect $\delta_{\wt{K}_3}$ acting on 
$\phi_{\g\ell} Y_{\ell m} e^{-i\o_{\g\ell} t}$ to effect $\ell \rightarrow \ell \pm 1$ without changing $m$ or the frequency. 
But because the frequency is proportional to $\g + \ell$ \eqref{gammaDef}, keeping the frequency unchanged while incrementing $\ell$ means 
$\g \rightarrow \g \mp 1$ at the same time. Explicit expressions for the effect of $\d_{\wt{K}_i}$, and the corresponding raising and lowering operators, are given in Appendix \ref{app:CKV_dS}.
Suffice to say the $\wt{K}_i$ in general mix up solutions with different $\ell$ and $m$ (while keeping frequency fixed) in a way reminiscent of how the $K_i$ act in the case of the black hole.

The two CKVs $\wt{K}_\pm$ introduce something more novel. They involve a factor of $e^{\pm t/L}$ and hence increment $i\o L$ by $1$. They do not involve $\theta, \varphi$ and thus do not affect $\ell, m$. They are therefore the analogs of $\d^{ij} \d_{P_i}\d_{P_j}$ and $\d^{ij} \d_{K_i}\d_{K_j}$ \eqref{eq:G_from_K} which raise and lower $\g$ (by $2$) without changing $\ell, m$.
A crucial difference is that here there is no need to form quadratic combinations of generators, and so $\wt{K}_\pm$ can increment $\g$ by $1$:
\begin{align}
\d_{\wt{K}_\pm}\left(\f_{\g\ell} Y_{\ell m} e^{-i\o_{\g\ell} t}\right) 
&= 
\frac{1}{L}\wt{G}^{\pm}_{\g\ell} \f_{\g\ell} Y_{\ell m} e^{-(i\o_{\g\ell}L\pm1)t/L},
\label{K+-}
\end{align}
where
\begin{align}
\wt{G}^+_{\g\ell} &\equiv -\sqrt{\D}\del_r 
- \frac{\sqrt{\D}}{r} 
- i \o_{\g\ell} L \frac{r}{\sqrt{\D}}\\
\wt{G}^-_{\g\ell} &\equiv \sqrt{\D}\del_r 
+ \frac{\sqrt{\D}}{r} 
- i \o_{\g\ell} L \frac{r}{\sqrt{\D}},
\end{align}
with $\wt{G}^\pm_{\g\ell}$ effecting $\g\pm1$. These ladder operators are equivalent to those in Section 3.1 of \cite{Anninos:2011af}. 
The various identities obeyed by $\wt{G}_{\g\ell}^\pm$, showing explicitly how they effect changes to the equation of motion operator, are in Appendix \ref{app:ladder_identities}.
As before, the corresponding contiguous relations for hypergeometric functions are shown in Appendix \ref{app:hyper_id}.
And as before, an economical way to build the entire (non-unitary) representation of the $\fr{so}(2,4)$ is to use the traceless symmetric tensor from \eqref{cYellm}
to recast one of our ladders:
\begin{align}
c^{\ell m}_{i_1\cdots i_\ell} 
\d_{\wt{K}_{i_1}} \cdots \d_{\wt{K}_{i_\ell}} 
\F_{\g 00}(t,r,\th,\varphi)
&\propto \F_{\g-\ell,\ell m}(t,r,\th,\varphi).
\end{align}
The proof is given in Appendix \ref{app:identities}. Recasting 
\eqref{K+-} in the same language, we have
\be
\d_{\wt{K}_\pm} \Phi_{\g\ell m} \propto \Phi_{\g\pm 1 , \ell m} .
\ee

The derivation of the quasinormal spectrum given in the last subsection is general, and makes no assumption about the value of the coupling to Ricci. Thus \eqref{eq:QNM_freq}, namely $i \o_{\g\ell} L = \g + \ell = \g_{\pm} + 2n + \ell$, should hold true for conformal coupling as well. What might seem puzzling is the fact that this spectrum has $\g$ (or $i\o_{\g\ell} L$) incrementing by $2$, while as argued earlier, $\wt{K}_{\pm}$ increments $\g$ by $1$. The two statements are in fact consistent, since for conformal coupling, $\g_- = 1$ and $\g_+ = 2$. In other words, one spectrum goes as $\g = 1, 3, 5, \ldots$, and the other spectrum goes as $\g = 2, 4, 6, \ldots$. From the point of view of the isometry algebra $\fr{so}(1,4)$, these form two separate representations. But viewed through the lens of the larger algebra $\fr{so}(2,4)$, the two together form a single irreducible representation, thanks to $\wt{K}_{\pm}$ which effects $\gamma \rightarrow \gamma \pm 1$.


Let us close this section with the remark that there are analogs of horizontal symmetries as in the black hole case \eqref{eq:Q0H0} \eqref{eq:QellHell} (one for each ($\g, \ell$)), from which conserved charges can be deduced. They can also be used to derive the quasinormal spectrum, and to explain the single-asymptote behavior of quasinormal modes. Explicit expressions for the symmetries and charges are given in Appendix \ref{app:ladder_identities}.

\section{Discussion}
\label{discussion}

In this note we have adopted a geometric point of view that is applicable to spin $0$ perturbations around both the black hole and de Sitter space. For the black hole, static, scalar perturbations enjoy an exact SO(1,3) symmetry, whose generators consist of 3 rotational KVs and 3 ``boost'' CKVs. The latter might come as a surprise since the scalar in question is not conformally coupled. But a special condition we call melodic \eqref{melodicDef} is responsible for the relevance of the 3 ``boost'' CKVs.\footnote{It is not surprising that the effective 3D metric \eqref{eq:ds3D} seen by the static scalar is conformally flat, and thus has CKVs. What is non-trivial is that some of its CKVs are melodic.} For de Sitter space, the dynamical scalar perturbations enjoy an exact SO(1,4) symmetry following from the isometry, which is extended to SO(2,4) if the scalar is conformally coupled. 

In both the case of the black hole, and that of de Sitter space, the perturbation solutions form representations of the corresponding algebra. The non-trivial algebra means the matrix representations of some symmetry generators must have non-vanishing off-diagonal terms. This is the origin of ladder structures. The most familiar example is the algebra of rotation generators $J_1, J_2, J_3$: in a basis in which $J_3$ is diagonal (i.e. with the solutions labeled by $m$), $J_1$ and $J_2$ are represented by non-diagonal matrices which means they give rise to ladder operators that can be used to raise or lower $m$. Static, scalar perturbations around the black hole can be labeled by $\ell, m$ (the quantum numbers of rotations), forming a unitary (principal series) representation of SO(1,3): the boost CKVs, which have a non-trivial algebra with rotations, effect transformations that raise or lower $\ell$. Dynamical scalar perturbations around de Sitter space are labeled by $\ell, m$ and frequency, forming a non-unitary representation of SO(1,4) (or SO(2,4) in the case of conformal coupling). Those symmetry generators that do not commute with the time-translation KV, effect transformations that raise or lower the frequency.

For both the black hole and de Sitter space, the nature of the perturbation solution in the ``ground state'', from which the whole representation can be generated by applying ladder operators, is what lies behind the remarkable single-asymptote property: i.e., the black hole perturbation solution that is regular at the horizon has purely ``growing'' tidal behavior but no ``decaying'' (Love number) tail at infinity; the quasinormal solution around de Sitter space is regular at the origin and purely outgoing at the horizon.

There are a number of interesting follow-up questions. How could this geometric/ representation-theoretic understanding be extended to spin 1 and spin 2 perturbations? For the black hole, hints of this can be gleaned from the spin ladder discussed in \cite{Hui:2021vcv}, where the spin ladder was used to explain the vanishing of spin 1 and 2 Love numbers for Kerr black holes. How could the geometric/representation-theoretic understanding be extended to {\it non-static} perturbations around a black hole? Near-zone approximations exist, which are applicable to low frequency black hole perturbations. One of them has a larger symmetry group than SO(1,3), in fact SO(2,4) \cite{Hui:2022vbh}. This might be useful for understanding the dissipative tidal response of the black hole. More ambitiously, to understand black hole quasinormal modes, one must go beyond the low frequencies associated with near-zone approximations. An interesting recent development is the identification of symmetries associated with a near-light-ring approximation \cite{Raffaelli:2021gzh, Hadar:2022xag, Kapec:2022dvc}. Could there be further hidden {\it exact} symmetries governing dynamical black hole perturbations? We hope to address some of these questions in the future.

We are grateful to Frederik Denef, Manvir Grewal, Klaas Parmentier, Alessandro Podo, and Luca Santoni for very helpful discussions and comments, and to collaborators Austin Joyce, Riccardo Penco, Luca Santoni, and Adam Solomon for many useful insights. 
Research for this work was supported in part by a Simons Fellowship in Theoretical Physics and the Department of Energy DE-SC011941. ZS is supported by the US National Science Foundation under Grant No. PHY-2209997 and the Gravity Initiative at
Princeton University.

\appendix

\section{(C)KVs of the 3D Effective Metric for Black Hole}
\label{app:CKV_S3}

The three KVs and three melodic CKVs of the 3D effective metric for the black hole are listed in \eqref{eq:BH_KV} and \eqref{eq:BH_MCKV}. The four non-melodic CKVs are
\begin{align}
\wt{K}_1^i \del_i &= 
\frac{1}{2} \D'(r) \sqrt{\D(r)} \sin\th \cos\varphi \, \del_r 
+ \sqrt{\D(r)}
\left(-\cos\th\cos\varphi \, \del_\th
+ \csc\th\sin\varphi \, \del_\varphi\right) \nonumber \\
\wt{K}_2^i \del_i &= 
\frac{1}{2} \D'(r) \sqrt{\D(r)} \sin\th \sin\varphi \, \del_r
- \sqrt{\D(r)}
\left(\cos\th\sin\varphi \, \del_\th
+ \csc\th\cos\varphi \, \del_\varphi\right) \nonumber \\
\wt{K}_3^i \del_i&= 
\frac{1}{2} \D'(r) \sqrt{\D(r)} \cos\th \, \del_r 
+ \sqrt{\D(r)}\sin\th \, \del_\th \nonumber \\
\wt{D}^i \del_i &= \sqrt{\D(r)} \del_r.
\end{align}
The commutation relations obeyed by the (C)KVs are
\begin{align}
\left[J_i, J_j\right] &= -\e_{ijk} J_k 
& 
\left[K_i, K_j\right] &= \b^2\e_{ijk} J_k 
&
\left[\wt{K}_i, \wt{K}_j\right] &= -\b^2\e_{ijk} J_k
\nonumber\\
\left[J_i, K_j\right] &= -\e_{ijk} K_k 
&
\left[J_i, \wt{K}_j\right] &= -\e_{ijk} \wt{K}_k 
&
\left[K_i, \wt{K}_j\right] &= -\b^2 \d_{ij}\wt{D} 
\nonumber\\
\left[\wt{D}, J_i\right] &= 0
&
\left[\wt{D}, K_i\right] &= \wt{K}_i
& 
\left[\wt{D}, \wt{K}_i\right] &= K_i, 
\end{align}
where $\b = \frac{1}{2}\left(r_+ - r_-\right).$ By taking
\be M_{ij} = \e_{ijk} J_k, \quad
M_{0i} = \frac{1}{\b} K_i, \quad
M_{i4} = -\frac{1}{\b}\wt{K}_i, \quad
M_{04} = \wt{D},\ee
we can show that these CKVs form an $\fr{so}(1,4)$ algebra. There are several interesting subalgebras. The $J_i$ form the isometry algebra $\fr{so}(3)$. The $K_i$ are melodic CKVs; together with the $J_i$, they form the melodic conformal algebra $\fr{so}(1,3)$.

\section{(C)KVs of $\text{dS}_4$}
\label{app:CKV_dS}

The 10 KVs and the five CKVs of de Sitter space are listed in \eqref{eq:dS_KV} and \eqref{eq:dS_CKV}. The commutation relations obeyed by the (C)KVs are
\begin{align}
\left[J_i, J_j\right] &= -\e_{ijk} J_k 
& 
\left[P_i, P_j\right] &= 0
&
\left[K_i, K_j\right] &= 0\nonumber\\
\left[\wt{K}_i, \wt{K}_j\right] 
&= \frac{1}{L^2}\e_{ijk}J_k
&
\left[J_i, P_j\right] &= -\e_{ijk}P_k
&
\left[J_i, K_j\right] &= -\e_{ijk}K_k
\nonumber\\
\left[\wt{K}_i, P_j\right] 
&= -\d_{ij}\wt{K}_+
&
\left[\wt{K}_i, K_j\right] 
&= \d_{ij}\wt{K}_- 
&
\left[J_i, \wt{K}_j\right] 
&= \e_{ijk}\wt{K}_k
\nonumber\\
\left[K_i, P_j\right] 
&= 2\d_{ij}D - 2 \e_{ijk}J_k
& 
\left[\wt{K}_+, P_j\right] &= 0 
&
\left[\wt{K}_-, P_j\right] &= -2 L \wt{K}_i
\nonumber\\
\left[\wt{K}_+, K_i\right] &= 2 \wt{K}_i
&
\left[\wt{K}_-, K_i\right] &= 0 
&
\left[\wt{K}_+, \wt{K}_i\right] &= \frac{1}{L^2} P_i
\nonumber\\
\left[\wt{K}_-, \wt{K}_i\right] &= -\frac{1}{L^2} K_i
&
\left[\wt{K}_\pm, J_i\right] &= 0
&
\left[\wt{K}_+, \wt{K}_-\right] &= -\frac{2}{L^2} D
\nonumber\\
\left[D, J_i\right] &= 0
&
\left[D, P_i\right] &= P_i 
&
\left[D, K_i\right] &= -K_i
\nonumber\\ 
\left[D, \wt{K}_i\right] &= 0 
& 
\left[D, \wt{K}_\pm\right] &= \pm \wt{K}_\pm.
& 
\end{align}
We saw in \eqref{eq:dS_to_M} the change of basis that turns the KVs into the standard basis of $\fr{so}(1,4)$. To this we add
\be
M_{05} = \frac{L}{2} 
\left(\wt{K}_+ + \wt{K}_-\right), \quad
M_{i5} = L\wt{K}_i, \quad
M_{45} = -\frac{L}{2}
\left(\wt{K}_+ - \wt{K}_-\right), \label{eq:dS_to_M2} 
\ee
which shows that the (C)KVs form the conformal algebra $\fr{so}(2,4)$, where we take $\h \equiv \text{diag}(-1,1,1,1,1,-1)$.

Once again, there are several interesting subalgebras. The $J_i$ again form an $\fr{so}(3)$ algebra. The $J_i$ and $\wt{K}_i$ together form an $\fr{so}(1,3)$ algebra. The four sets $\{D, \wt{K}_+, \wt{K}_-\}$ and $\{D, K_i, P_i\}$ for any $i$ form $\fr{sl}\left(2,\mathbb{R}\right) \cong \fr{so}(1,2)$ algebras. 

The fact that $J_i$ and $\wt{K}_i$ form an $\fr{so}(1,3)$ algebra is similar to the black hole case. For instance, the analog of \eqref{eq:RN_delta_K3} is
\begin{align}
\d_{\wt{K}_3}\left(\f_{\g\ell} Y_{\ell m} e^{-i\o_{\g\ell} t}\right) 
&= e^{-i\o_{\g\ell} t}
\left(f(\ell) \wt{D}^+_{\ell} \f_{\g\ell} Y_{\ell+1,m}
- f(\ell-1) \wt{D}^-_{\ell} \f_{\g\ell} Y_{\ell-1,m}\right),
\label{eq:dS_conf_delta_K3}
\end{align}
where 
\begin{align}
\label{tildeDdef}
\wt{D}^+_{\ell} &\equiv \left(1-\frac{r^2}{L^2}\right)\del_r 
- \frac{r}{L^2} - \frac{\ell}{r} \nonumber \\
\wt{D}^-_{\ell} &\equiv -\left(1-\frac{r^2}{L^2}\right)
\del_r + \frac{r}{L^2} - \frac{\ell+1}{r}.
\end{align}
We see that $\wt{D}^+_{\ell}$ effect $\ell \pm 1$ while keeping $\o$ fixed. However, when indexing the solutions with $\g$, holding $\o$ fixed while raising/lowering $\ell$ means lowering/raising $\g$ to compensate. Thus the $\o_{\g\ell}$ factor in the exponent in \eqref{eq:dS_conf_delta_K3} should really be considered $\o_{\g - 1,\ell+1}$ when multiplying the first term in parentheses and $\o_{\g+1,\ell-1}$ when multiplying the second term, though of course these both equal $\o_{\g\ell}$.

\section{Solutions to Equations of Motion}
\label{app:solutions}

For the Reissner--Nordstr\"{o}m black hole, the two independent solutions to \eqref{eq:RN_EoM} are
\begin{align}
\f_{\ell}^{(1)}(r) 
&=
{}_2 F_1\left(-\ell, \ell+1, 1, x\right)
\label{eq:RN_sol1}
\\
\begin{split}
\label{eq:RN_sol2}
\f_{\ell}^{(2)}(r) 
&= 
{}_2 F_1\left(-\ell, \ell+1, 1, x\right)
\log\left(\frac{x}{x-1}\right)
\\
&\quad + 
\sum_{k=0}^{\ell-1}
\left[(-1)^k \binom{\ell+k}{k}\binom{\ell}{k}
\left(\psi(\ell-k+1) + \psi(\ell+k+1) - 2 \psi(k+1)\right)x^k
\right.\\
&\quad
\left.
- \left(\sum_{j=0}^{k-1} (-1)^j \binom{\ell+j}{j} 
\binom{\ell}{j} \frac{1}{k-j}\right) x^k
\right],
\end{split}
\end{align}
where $x = \frac{r - r_-}{r_+ - r_-}$. At large $r$, the first goes at $r^\ell$, and the second goes as $1/r^{\ell+1}$. 

For four-dimensional de Sitter space, the two independent solutions to \eqref{eq:eomdS} are
\begin{align}
\f^{(1)}_{\g\ell}(r) &= 
\left(\frac{r^2}{L^2}\right)^{\ell/2} \left(1-\frac{r^2}{L^2}\right)^{\frac{i\o_{\g\ell} L}{2}} \, 
{}_2F_1\left(a,b,c;\frac{r^2}{L^2}\right)
\label{eq:dS_sol1}
\\
\f^{(2)}_{\g\ell}(r) &= 
\left(\frac{r^2}{L^2}\right)^{-(\ell+1)/2} \left(1-\frac{r^2}{L^2}\right)^{\frac{i\o_{\g\ell} L}{2}} \, 
{}_2F_1\left(a-c+1,b-c+1,2-c;\frac{r^2}{L^2}\right),
\label{eq:dS_sol2}
\end{align}
where 
\be a = \frac{1}{2}\left(\frac{3}{2} + \ell + i\o_{\g\ell} L + \sqrt{\a}\right), \quad
b = \frac{1}{2}\left(\frac{3}{2} + \ell + i\o_{\g\ell} L - \sqrt{\a}\right), \quad
c = \ell + \frac{3}{2}, 
\label{eq:dS_param}
\ee
with $\a = \frac{9}{4} - 12\x$.

Near the origin, the first goes at $r^\ell$, and the second goes as $1/r^{\ell+1}$.

\section{Hypergeometric Identities}
\label{app:hyper_id}

The hypergeometric differential equation is
\be z(1-z)\frac{d^2}{dz^2}F(z) 
+ (c-(a+b+1)z)\frac{d}{dz}F(z) - a b F(z) = 0,\ee
and is solved by the two functions
\be {}_2F_1(a,b,c;z), \quad
z^{1-c} {}_2F_1(a-c+1,b-c+1,2-c;z), \nonumber \ee
where
\be 
{}_2F_1(a,b,c;z) = \sum_{n=0}^\infty 
\frac{(a)_n (b)_n}{(c)_n} \frac{z^n}{n!}.\ee
Note the symmetry in $a$ and $b$, which will be useful to us. When $c$ is a non-positive integer, the second solution is instead given by the first times $\log(z)$, plus a different power series. This is relevant for the Reissner--Nordstr\"{o}m case, as we can see in \eqref{eq:RN_sol2} above. 

Hypergeometric functions with the parameters $a$ $b$, or $c$ offset by 1 can be related to the original function via the 15 contiguous relations, which relate ${}_2F_1(a,b,c;z)$ to any two of the six functions
\be {}_2F_1(a\pm1,b,c;z), \quad 
{}_2F_1(a,b\pm1,c;z), \quad 
{}_2F_1(a,b,c\pm1;z). \ee
Repeated applications of these relations can yield a relation between any ${}_2F_1(a,b,c;z)$ and any two hypergeometric functions with parameters offset by any three integers. We shall discuss the identities used to uncover the various ladders here. For an extensive discussion of hypergeometric functions and their various identities, see \cite{Magnus, ContigRakha, Stalker}. 

The $D^\pm_\ell$ operators raise and lower $\ell$ for the static solutions in the Reissner--Nordstr\"{o}m background. We see from \eqref{eq:RN_sol1} that raising $\ell$ corresponds to lowering $a$ and raising $b$, and vice versa for lowering $\ell$. Thus the ladder arises directly from these two identities:
\small
\begin{align} 
\left((b-a-1)z(1-z)\frac{d}{dz} 
- a(c-b + (b-a-1)z)\right) {}_2F_1(a,b,c;z)
&= a (c-b) \, {}_2F_1(a+1,b-1,c;z)\\
\left((a-b-1)z(1-z)\frac{d}{dz} 
- b(c-a - (a-b-1)z)\right) {}_2F_1(a,b,c;z)
&= b (c-a) \, {}_2F_1(a-1,b+1,c;z).
\end{align} 
\normalsize

The $E^\pm_{\g\ell}$ operators raise and lower $\ell$ while keeping $\g$ fixed for the solutions in de Sitter. Examining \eqref{eq:dS_sol1} and \eqref{eq:dS_param}, we see that this entails raising and lowering $a$, $b$, and $c$ all at once. Thus we need the identities
\begin{align} 
\frac{d}{dz} \, {}_2F_1(a,b,c;z)
&= \frac{ab}{c} \, {}_2F_1(a+1,b+1,c+1;z) \\ 
\frac{d}{dz} \left(z^{c-1} (1-z)^{a+b-c} 
\, {}_2F_1(a,b,c;z)\right)
&= (c-1) z^{c-2} (1-z)^{a+b-c-1} 
\, {}_2F_1(a-1,b-1,c-1;z).
\end{align}
Of course, because the de Sitter equation of motion requires a change of variables to $z = r^2/L^2$ and a field redefinition to turn bring it into hypergeometric form, the exact form of the ladder operators is not immediately apparent from these identities. 

The $F^\pm_{\g\ell}$ operators raise and lower $\g$ while simultaneously lowering and raising $\ell$ and thus require identities which keep $a$ and $b$ fixed while raising and lowering $c$. These are easily supplied:
\begin{align} 
\frac{d}{dz} \left((1-z)^{a+b-c} 
\, {}_2F_1(a,b,c;z)\right)
&= \frac{(c-a)(c-b)}{c} (1-z)^{a+b-c-1} 
\, {}_2F_1(a,b,c+1;z) \\ 
\frac{d}{dz} \left(z^{c-1} 
\, {}_2F_1(a,b,c;z)\right)
&= (c-1) z^{c-2} 
\, {}_2F_1(a,b,c-1;z) 
\end{align}

The $G^\pm_{\g\ell}$ operators, which raise and lower $\g$ by 2 while keeping $\ell$ fixed, can be written in terms of previous two, but also arise from identities which raise and lower both $a$ and $b$ while keeping $c$ fixed:
\begin{align} 
\left((A+1) z \frac{d}{dz} + a b \right)
\, {}_2F_1(a,b,c;z)
&= a b (1-z) \, {}_2F_1(a+1,b+1,c;z)\\
\left((A-1) z (1-z) \frac{d}{dz} 
- \left(B - C(1-z)\right)\right)
{}_2F_1(a,b,c;z)
&= -(c-a)(c-b) \, {}_2F_1(a-1,b-1,c;z),
\end{align}
where
\be A = a+b-c, \quad
B = A(A-1), \quad
C = a(A-1) + (b-1)(b-c). \nonumber \ee

Moving to the conformally coupled case, the $\wt{D}^\pm_{\ell}$ operators raise and lower $\ell$ while simultaneously lowering and raising $\g$ to keep $\o$ fixed. This would seem to indicate that we need an identity that raises and lowers $a$ and $b$ by $1/2$ and $c$ by 1. To our knowledge, no such identity exists. However, since the hypergeometric function is symmetric in $a$ and $b$ and in the conformally coupled case $a = b+1/2$, raising and lowering both $a$ and $b$ by $1/2$ is equivalent to keeping $a$ fixed and raising $b$ by 1 and lowering $a$ by 1 and keeping $b$ fixed, respectively. Thus the identities we require are
\begin{align} 
\frac{d}{dz} \left((1-z)^{b} 
\, {}_2F_1(a,b,c;z)\right)
&= \frac{b(a-c)}{c} (1-z)^{b-1} 
\, {}_2F_1(a,b+1,c+1;z) \\ 
\frac{d}{dz} \left(z^{c-1} (1-z)^{b-c+1} 
\, {}_2F_1(a,b,c;z)\right)
&= (c-1) z^{c-2} (1-z)^{b-c} 
\, {}_2F_1(a-1,b,c-1;z).
\end{align}

Lastly, the $\wt{G}^\pm_{\g\ell}$ operators raise and lower $\g$ while keeping $\ell$ fixed. This seems to requires raising and lowering both $a$ and $b$ by 1/2, but by the same logic as before, we can instead use the following identities:
\begin{align} 
\frac{d}{dz} \left(z^b \, {}_2F_1(a,b,c;z)\right)
&= b z^{b-1} \, {}_2F_1(a,b+1,c;z) \\ 
\frac{d}{dz} \left(z^{c-a} (1-z)^{a+b-c} \, {}_2F_1(a,b,c;z)\right)
&= (c-a) z^{c-a-1} (1-z)^{a+b-c-1} \, {}_2F_1(a-1,b,c;z).
\end{align}

\section{Commutators and Lie Brackets}
\label{app:commutator}

In this paper, when we write something like
$[X, Y]=Z$ where $X, Y, Z$ are vectors, we implicitly mean three different, mutually consistent statements.
The first is interpreting $[X, Y]$ as a Lie bracket \eqref{LieBracket}, i.e.
$[X, Y]^\mu_{\rm LB} = Z^\mu$. 
The second is interpreting it as a statement about Lie derivatives: $[\La_X , \La_Y] = \La_Z$ \eqref{LieDerivCom}. 
The third is interpreting it as a a statement about symmetry
transformation \eqref{dXLieX}: $[\delta_X , \delta_Y] = \delta_Z$ \eqref{deltaXdeltaY1}. The goal of this Appendix is to demonstrate their mutual consistency.

In this paper, we are interested in symmetry formation of the form
\be\label{dXLieX} \d_X = \La_X + c \left(\nabla_\m X^\m\right),\ee
where $X$ any vector, $c$ is any constant, and $\La_X$ denotes the Lie derivative along $X$, obey the same algebra as the Lie derivatives along the vectors themselves when acting on a general tensor. Here (and only here) we shall carefully distinguish between commutators of differential operators and Lie brackets of vectors, denoting the latter by
\be 
\label{LieBracket}
\left[X,Y\right]_\text{LB}^\m 
\equiv X^\m \nabla_\m Y^\n 
- Y^\m \nabla_\m X^\n.
\ee
We then cite the well-known property that for any tensor $T$ of rank $(r,s)$,
\be 
\label{LieDerivCom}
\left[\La_X, \La_Y\right]
{T^{\m_1 \cdots \m_r}}_{\n_1\cdots\n_s} = \La_{\left[X,Y\right]_\text{LB}}
{T^{\m_1 \cdots \m_r}}_{\n_1\cdots\n_s} , \ee
and compute
\begin{align}
\label{deltaXdeltaY}
\left[\d_X, \d_Y\right]
{T^{\m_1 \cdots \m_r}}_{\n_1\cdots\n_s}
&= \big(\left[\La_X, \La_Y\right]
+ c\left(\La_X\left(\nabla_\m Y^\m\big)
- \La_Y\left(\nabla_\m X^\m\right)\right)\right)
{T^{\m_1 \cdots \m_r}}_{\n_1\cdots\n_s}\nonumber\\
&= \left(\La_{\left[X,Y\right]_\text{LB}}
+ c\left(X^\n \nabla_\n \nabla_\m Y^\m
- Y^\n \nabla_\n\nabla_\m X^\m\right)\right)
{T^{\m_1 \cdots \m_r}}_{\n_1\cdots\n_s}~.
\end{align}
In the second line of (\ref{deltaXdeltaY}), replacing the double covariant derivative $\nabla_\nu\nabla_\mu$ by $\nabla_\mu\nabla_\nu-R_{\mu\nu}$ allows us to express $X^\n \nabla_\n \nabla_\m Y^\m
- Y^\n \nabla_\n\nabla_\m X^\m$ as the divergence of $[X, Y]_{\rm LB}$, and hence we find 
\begin{align}\label{deltaXdeltaY1}
\left[\d_X, \d_Y\right]
{T^{\m_1 \cdots \m_r}}_{\n_1\cdots\n_s} &= \d_{\left[X,Y\right]_\text{LB}}
{T^{\m_1 \cdots \m_r}}_{\n_1\cdots\n_s}.
\end{align}
Thus we see that any symmetry transformations of the form above obey the same algebra as the corresponding vectors.

\section{Melodic Conformal Killing Vectors}
\label{app:MCKV}

We consider a $d$-dimensional pseudo-Riemannian manifold $\mc{M}_d$ with metric $g_{\m\n}$. A CKV of this manifold is a vector $X$ which satisfies
\be
\nabla^\m X^\n + \nabla^\n X^\m
= \frac{2}{d}\nabla_\r X^\r g^{\m\n}.
\label{eq:CKE}
\ee
An ordinary KV satisfies the above equation with $\nabla_\r X^\r = 0$. Taking derivatives of the conformal Killing equation yields the following identities, which shall prove useful later:
\begin{align}
\Box X^\m &= -\frac{d-2}{d} \nabla^\m \nabla_\n X^\n - {R^\m}_\n X^\n
\label{eq:box_id}
\\
\Box \nabla_\m X^\m
&= -\frac{1}{d-1}\left(R \nabla_\m X^\m + \frac{d}{2} X^\m \nabla_\m R\right)
\label{eq:box_nabla_id} \nonumber\\
&= d \, \nabla_\m \Box X^\m.
\end{align}
CKVs naturally arise in the context of conformal field theories. As shown in the inset in Section \ref{BHgeom}, a CKV has a naturally action on scalars that is a symmetry of the action for conformal coupling, but more surprisingly, for a generic Ricci coupling, a melodic CKV, that is, one obeying
\be \Box \nabla_\m X^\m = 0,\ee
also yields a symmetry of the action.\footnote{If we include a mass term as well, the condition instead becomes $(d-1)\left(\x-\x_c\right)\Box\nabla_\m X^\m - m^2 \nabla_\m X^\m = 0$.}

A few properties of melodic CKVs are quickly apparent from \eqref{eq:box_nabla_id}. If the manifold has $R = 0$, all CKVs are melodic. If the manifold is maximally symmetric but not flat, so $\nabla_\m R = 0$ but $R \neq 0$, then a CKV is melodic if and only if $\nabla_\m X^\m = 0$, which means $X$ is just an ordinary KV. It can be shown that if $R \neq 0$ and $X$ is melodic CKV, a rescaling of the metric $g_{\m\n} \to \O^2 g_{\m\n}$ will make $X$ a KV of the new metric if $\O^2 = R L_0^2$, where $L_0$ is an arbitrary length scale.

An important property is that the set of melodic CKVs is closed under Lie brackets, which we can verify with the following identity that holds for any two CKVs $X$ and $Y$:
\be
\begin{split}
\Box \nabla_\n \left(X^\m \nabla_\m Y^\n - Y^\m \nabla_\m X^\n\right)
&= X^\m \nabla_\m \Box\nabla_\n Y^\n
- Y^\m \nabla_\m \Box\nabla_\n X^\n\\
&\quad+ \frac{2}{d} \nabla_\m X^\m 
\Box\nabla_\n Y^\n
- \frac{2}{d} \nabla_\m Y^\m 
\Box\nabla_\n X^\n.
\end{split}
\ee
Thus if both $X$ and $Y$ are melodic CKVs, so is $[X,Y]$. 

It follows that the melodic CKVs form a Lie algebra. Since the KVs form the isometry algebra of the manifold, and the CKVs form the conformal algebra, the algebra formed by the melodic CKVs must lie ``in between'' these two algebras. Denoting this algebra with $\fr{mconf}\left(\mc{M}_d\right)$ and letting 
$\fr{isom}\left(\mc{M}_d\right)$ and $\fr{conf}(\mc{M}_d)$ be the isometry and conformal algebras, respectively, we have the following hierarchy:
\be
\fr{isom}\left(\mc{M}_d\right)
\subset \fr{mconf}\left(\mc{M}_d\right)
\subset \fr{conf}\left(\mc{M}_d\right).
\ee
(Here we consider KVs to be trivially melodic CKVs.) 
We see from above that when $\mc{M}_d$ is Ricci-scalar-flat ($R = 0$), $\fr{mconf}\left(\mc{M}_d\right)= \fr{conf}\left(\mc{M}_d\right)$ because all CKVs are melodic, and when $\mc{M}_d$ is maximally symmetric but not flat, $\fr{isom}\left(\mc{M}_d\right) = \fr{mconf}\left(\mc{M}_d\right)$, as the only melodic CKVs are ordinary KVs. Outside of these extreme cases, the algebra $\fr{mconf}\left(\mc{M}_d\right)$ is somewhat mysterious. The three-dimensional manifold that appeared in Section \ref{BHgeom} as the effective space for static scalars in a Reissner–Nordstr\"{o}m background is the only space we have found with a non-trivial algebra of melodic CKVs. Letting that space be $\S_3$, we have
\be
\fr{isom}\left(\S_3\right) = \fr{so}(3)
\subset \fr{mconf}\left(\S_3\right) = \fr{so}(1,3)
\subset \fr{conf}\left(\S_3\right) = \fr{so}(1,4).
\ee

Returning to a scalar field theory, we are naturally led to consider the conserved currents of the symmetries associated with the melodic CKVs. The Noether procedure yields
\be J_\m = T_{\m\n}X^\n + Z_\m,
\ee
where 
\be Z_\m \equiv \frac{2(d-1)}{d}
\left(\x-\x_c\right)\left(\frac{1}{2}\F^2 \nabla_\m\nabla_\n X^\n 
- \F \nabla_\m \F \nabla_\n X^\n\right),\ee
and the conserved stress-energy tensor is 
\be T_{\m\n} = \nabla_\m \F \nabla_\n \F
- \frac{1}{2} g_{\m\n} \nabla_\r \F \nabla^\r \F
+ \x \F^2 G_{\m\n} 
+ \x\left(g_{\m\n} \nabla_\r \nabla^\r 
- \nabla_\m \nabla_\n\right)\left(\F^2\right).\ee
The trace of the stress energy tensor is
\be {T^\m}_\m = (d-1)\left(\x - \x_c\right)
\nabla_\m \left(\F^2\right)
- \frac{d-2}{2} \F\left(\Box - \x R\right)\F.\ee
Thus we see that for $\x = \x_c$, the stress-energy tensor is traceless on the equation of motion, which is a well-known property of conformal field theories. It can be shown that 
\be \nabla_\m Z^\m = 
-\frac{1}{d}{T^\m}_\m \nabla_\n X^\n
+ \frac{d-1}{d} \F^2 \Box \nabla_\m X^\m
+ \frac{d-2}{2d} \nabla_\m X^\m 
\F\left(\Box - \x R\right)\F.
\label{eq:nablaZ}
\ee
The second term vanishes for a melodic CKV, and the third vanishes on the equation of motion. We can use \eqref{eq:nablaZ}, combined with the conservation of $T_{\m\n}$ on the equation of motion, the conformal Killing equation, and the melodic property of the CKV to show that the current we have defined is conserved on the equation of motion:
\begin{align} 
\nabla^\m J_\m &= \nabla^\m T_{\m\n} X^\n 
+ T_{\m\n} \nabla^\m X^\n 
+ \nabla^\m Z_\m\\
&= \frac{1}{d} {T^\m}_\m \nabla_\n X^\n + \nabla^\m Z_\m
+ \text{terms that vanish on-shell}\\
&= 0
+ \text{terms that vanish on-shell}.
\end{align}

\section{Ladder Identities in de Sitter space}
\label{app:ladder_identities}

In this appendix we collect several identities for the ladder operators 
in de Sitter space. The identities show in an explicit way how the ladder operators map a solution at one level to that at another level. 
We divide the discussion based on the coupling to Ricci.
We also provide the conserved horizontal charges for the conformally coupled case.

\noindent {\it Generic coupling.}
The equation of motion can be expressed as
\be
\label{Hgeneric}
H_{\g\ell} \phi_{\g\ell} = 0 \quad , \quad 
H_{\g\ell} \equiv
-\frac{1}{r^2}\left(\del_r\left(\D \del_r\right)
+ \frac{\o_{\g\ell}^2r^4}{\D} 
+ \left(\a - \frac{9}{4}\right)\frac{r^2}{L^2} 
- \ell(\ell+1) \right).
\ee 
For the $E_{\g\ell}^\pm$ and $F_{\g\ell}^\pm$ operators defined in
\eqref{EFdef}, we have the following ladder relations:
\be H_{\g,\ell+1} E_{\g\ell}^+ 
= E_{\g\ell}^+ H_{\g\ell},
\quad
H_{\g,\ell-1} E_{\g\ell}^- 
= E_{\g\ell}^- H_{\g\ell}.
\ee
\be 
H_{\g+2,\ell-1} F_{\g\ell}^+ 
= F_{\g\ell}^+ H_{\g\ell},
\quad
H_{\g-2,\ell+1} F_{\g\ell}^-
= F_{\g\ell}^- H_{\g\ell},
\ee
confirming what we saw from the geometric perspective in Section \ref{dSgeom}. 

We can then write
\begin{align}
&H_{\g\ell} 
= E_{\g,\ell-1}^+ E_{\g\ell}^- 
+ \frac{1}{L^2}a_{\g\ell}
= E_{\g,\ell+1}^- E_{\g\ell}^+ 
+ \frac{1}{L^2}a_{\g+2,\ell}
\label{eq:Eid}\\
&E^-_{\g,\ell+1} E^+_{\g\ell} 
- E^+_{\g,\ell-1} E^-_{\g\ell} 
= -\frac{2\left(2\g + 4\ell + 1\right)}{L^2}\\
&H_{\g\ell} 
= F_{\g-2,\ell+1}^+ F_{\g\ell}^- 
+ \frac{1}{L^2}b_{\g\ell}
= F_{\g+2,\ell-1}^- F_{\g\ell}^+ 
+ \frac{1}{L^2}b_{\g+2,\ell}
\label{eq:Fid}\\
&F^-_{\g+2,\ell-1} F^+_{\g\ell} 
- F^+_{\g-2,\ell+1} F^-_{\g\ell} 
= -\frac{2\left(2\g - 1\right)}{L^2},
\end{align}
where $a_{\g\ell}$ and $b_{\g\ell}$ are defined in \eqref{eq:const_ab}.

\noindent {\it Conformal coupling.}
For the conformally coupled scalar, to exhibit the ladder structure due to the CKV $\wt{K}_3$, it turns out to be convenient to write the equation of motion as
\be
H_{\g\ell} \phi_{\g\ell} = 0 \quad , \quad
H_{\g\ell} \equiv
-\frac{\D}{r^4}\left(\del_r\left(\D \del_r\right)
+ \frac{\o_{\g\ell}^2r^4}{\D} - 2\frac{r^2}{L^2} - \ell(\ell+1) \right).
\label{eq:H_Dtilde}
\ee 
This $H_{\g\ell}$ is proportional to the one in \eqref{Hgeneric}, with $\alpha = 1/4$.
For the ladder operators
$\wt{D}^\pm_\ell$ defined in \eqref{tildeDdef}, we have:
\be H_{\g-1,\ell+1} \wt{D}_\ell^+ = \wt{D}_\ell^+ H_{\g\ell}, \quad
H_{\g+1,\ell-1} \wt{D}_\ell^- = \wt{D}_\ell^- H_{\g\ell},
\label{eq:conf_ladder}
\ee
as well as
\begin{align} 
& H_{\g\ell} = \wt{D}_{\ell-1}^+ \wt{D}_\ell^- - \o_{\g\ell}^2
- \frac{\ell^2}{L^2} 
= \wt{D}_{\ell+1}^- \wt{D}_\ell^+ - \o_{\g\ell}^2
- \frac{(\ell+1)^2}{L^2}
\label{eq:Dtildeid}\\
&\wt{D}^-_{\ell + 1} \wt{D}^+_\ell - \wt{D}^+_{\ell - 1} \wt{D}^-_\ell 
= \frac{2\ell+1}{L^2}.
\end{align}
Observe that since $H_{\g+1,-1} = H_{\g 0}$ (recall that $\o_{\g\ell}$ is proportional to $\gamma + \ell$), we have 
\be 
\left[Q_0 , H_{\g0}\right] = 0 \quad {\rm where} \quad Q_0 \equiv \wt{D}^-_0 = -\frac{\D}{r^3}\partial_r r
\label{eq:Q0H0_conf}
\ee
and thus $\d\f_0 = Q_0\f_0$ is a symmetry. 
This can be verified at the level of the action as well. This is the horizontal symmetry at $\ell=0$, just as in the black hole case. Climbing the ladder, it can be shown 
\be 
\left[Q_\ell, H_{\g\ell}\right] = 0 \quad {\rm where} \quad Q_\ell \equiv 
\wt{D}^+_{\ell-1} Q_{\ell-1} \wt{D}^-_\ell,
\ee
and again $\d\f_\ell = Q_\ell \f_\ell$ can be shown to be a symmetry of the action. See Appendix \ref{app:horizontal} for a proof. 
For $\ell = 0$, the Noether procedure yields the conserved charge 
\be P_{\g0} \equiv \frac{\D^2}{r^4}\left(\del_r\left(r\f_{\g0}\right)\right)^2 
+ \o_{\g0}^2 r^2 \f_{\g0}^2 , 
\label{eq:P0}
\ee
in the sense that $\partial_r P_{\gamma 0} = 0.$
Note that the charge depends on $\g$, because of $\g$ dependence from both $\omega_{\g0}$ and $\phi_{\gamma 0}$/
We can again climb the ladder to define the charges at non-zero $\ell$:
\be P_{\g\ell} \equiv \frac{\D^2}{r^4}\left(\del_r
\left(r \wt{D}^-_1 \wt{D}^-_2 \cdots \wt{D}^-_\ell \f_{\g\ell}\right)\right)^2 
+ \o_{\g\ell}^2 r^2 \left(\wt{D}^-_1 \wt{D}^-_2 \cdots \wt{D}^-_\ell \f_{\g\ell}\right)^2 .
\label{eq:Pl}
\ee
It can be shown these are precisely the Noether charges of the symmetries effected by the $Q_\ell$ operators.\footnote{Since the $D^\pm_\ell$ operators change both $\ell$ and $\g$, we must take care with the subscripts here. The series of $\wt{D}^-_\ell$ operators acting on $\f_{\g\ell}$ produce $\f_{\g+\ell,0}$, and we can then use $\o_{\g\ell} = \o_{\g+\ell,0}$ and invoke conservation of $P_{\g+\ell,0}$ to show that $P_{\g\ell}$ is conserved.} Note that unlike in the black hole case, the conserved charges are not complete squares i.e. one cannot take square root to reduce them to be linear in the field.
If one constructs $\phi_{\g\ell}$ by
$\phi_{\g\ell} \propto \wt{D}^+_{\ell-1} ... \wt{D}^+_0 \phi_{\g+\ell , 0}$, 
it can be shown that
\be
P_{\g\ell} 
\propto \frac{1}{L^{4\ell}}
\left(\frac{\left(\g\right)_{2\ell+1}}{\g+\ell}\right)^2 P_{\g+\ell,0},\ee
where $(x)_n = \G(x+n)/\G(n)$ is the Pochhammer symbol.
Note that for $\g = -\ell$, the expression should be evaluated as a limit. Just as the conserved horizontal charges for black hole allowed us to show the vanishing of the Love numbers, the conservation of these charges gives us an alternative way to derive the spectrum of quasinormal mode frequencies in the conformally coupled case, which we shall not discuss here.

For the $\wt{G}_{\g\ell}^\pm$ operators associated with the CKVs $\wt{K}_\pm$ \eqref{K+-}, yet another choice of the equation of motion operator turns out to be useful:
\be \wt{H}_{\g\ell} \equiv \frac{r^4}{\D} H_{\g\ell} ,
\ee
for which we find the ladder relations:
\be
 \wt{H}_{\g+1,\ell} \wt{G}_{\g\ell}^+ = \wt{G}_{\g\ell}^+ \wt{H}_{\g\ell}, \quad
\wt{H}_{\g-1,\ell} \wt{G}_{\g\ell}^- = \wt{G}_{\g\ell}^- \wt{H}_{\g\ell},\ee
as well as
\begin{align}
&\wt{H}_{\g\ell} 
= \wt{G}^-_{\g+1,\ell} \wt{G}_{\g\ell}^+ 
- \g \left(\g + 2\ell + 1\right)
= \wt{G}_{\g-1,\ell}^+ \wt{G}_{\g,\ell}^- 
- \left(\g - 1\right)
\left(\g + 2\ell\right) 
\label{eq:conf_Gid}
\\
&\wt{G}^-_{\g+1,\ell} \wt{G}^+_{\g\ell} 
- \wt{G}^+_{\g-1,\ell} \wt{G}^-_{\g\ell} 
= 2(\g+\ell).
\end{align}

\section{Conditions for Horizontal Symmetries}
\label{app:horizontal}

In this appendix we shall establish the conditions for the existence of the so-called horizontal symmetries used throughout the main text. We consider an action of the form
\be S = \sum_n S_n = \frac{1}{2}\sum_n \int dr \, 
\f_n^* \mc{H}_n \f_n,
\label{eq:abstract_action}
\ee
where 
\be \mc{H}_n \equiv \del_r \D \del_r + V_n(r).
\label{eq:abstract_H}\ee
Here $\D(r)$ and $V_n(r)$ are arbitrary functions, and $n$ is an abstract index that can represent multiple indices. The equation of motion is naturally
\be \mc{H}_n \f_n = 0.\ee
We now assume that we have some ladder operators, i.e., some $\mc{D}^\pm_n$ that raise and lower $n$. We must have
\be \mc{D}^\pm_n H_n \f_n = H_{n\pm k} \mc{D}^\pm_n \f_n.
\label{eq:abstract_ladder}
\ee
where the equation of motion operator is given by
\be H_n \equiv -f(r) \mc{H}_n
\label{eq:ladder_vs_action}\ee
for some function $f(r)$, and we have introduced a minus sign for later convenience. Like $n$, $k$ is an abstract index that could represent the shifting of many different indices by different amounts. We may assume that the ladder operators are first order, since we may always remove higher derivatives by subtracting $H_n$ or its derivatives.

Lastly, for any differential operator $\mc{O}$, we let $\wh{\mc{O}}$ be the operator that arises from integrating by parts, i.e.,
if 
\be \mc{O} = \sum_i a_i(r) \del_r^i,\ee
then
\be \wh{\mc{O}} = \sum_i (-1)^i \del_r^i a_i(r),\ee
where derivatives act on everything to the right. We can then write
\be \int dr \, \f \mc{O} \psi 
= \int dr \, \psi \wh{\mc{O}} \f\ee
by dropping a boundary term. 

We now take $n = 0$ (or any specific value, but 0 is convenient) and consider some first order differential operator $Q_0$ which obeys $Q_0^* = Q_0$. We define the transformation $\d\f_0 = Q_0 \f_0$ and find that the action transforms as 
\begin{align}
\d S_0 &= \int dr \, \left(Q_0 \f_0^* \mc{H}_0\f_0 + \f_0^* \mc{H}_0 Q_0 \f_0\right)\\
&= \int dr \, \left(\f_0^*\left(\wh{Q}_0 \mc{H}_0 + \mc{H}_0 Q_0\right)\f_0\right).
\end{align}
Thus we see that a sufficient condition for $\d S_0$ to vanish is 
\be \wh{Q}_0 \mc{H}_0 = -\mc{H}_0 Q_0.
\label{eq:Q0_H_condition}
\ee
Since $Q_0$ is a first order operator, it can be written in the form 
\be Q_0 = a(r) \del_r b(r)\ee
for some functions $a(r)$ and $b(r)$. It follows that $\wh{Q}_0$ can be written
\be \wh{Q}_0 = -b(r) \del_r a(r) = -\frac{1}{g(r)} Q_0 g(r)
\label{eq:Q0hat},\ee
where $g(r) = a(r)/b(r)$. If
\be \left[Q_0, g(r) \mc{H}_0\right] = 0,
\label{eq:Q0_H_commutator}
\ee
we can write
\be
\wh{Q}_0\mc{H}_0 = -\frac{1}{g(r)} Q_0 g(r) \mc{H}_0
= -\mc{H}_0 Q_0,
\ee
so \eqref{eq:Q0_H_condition} is met. Thus if we can find some operator $Q_0$ that commutes with $g(r) \mc{H}_0$ for the same $g(r)$ that appears in \eqref{eq:Q0hat}, then $\d\f_0 = Q_0 \f_0$ is a symmetry of the action $S_0$. 

It is worth pausing to note that at this stage, the ladder structure has not been used at all, and as noted previously, $n$ is set to zero simply for concreteness. If we somehow encounter any operator $Q_n$ obeying \eqref{eq:Q0hat} and \eqref{eq:Q0_H_commutator}, then we have an associated symmetry of the action $S_n$. Having a ladder structure may make it easier to find such a $Q_n$ but is in no way necessary.

If we have such an operator $Q_0$, we can then define
\be Q_n = \mc{D}^+_{n-k} Q_{n-k} \mc{D}^-_n,
\label{eq:abstract_Q_recursion}
\ee
assuming $n$ and $k$ are such that this recursion terminates at $Q_0$. We also assume $\left(\mc{D}^\pm_n\right)^* = \mc{D}^\pm_n$. Under what circumstances is $\d\f_n = Q_n \f_n$ a symmetry of the action $S_n$? The same procedure as in the $n = 0$ case leads us to the sufficient condition
\be \wh{Q}_n\mc{H}_n = -\mc{H}_n Q_n.
\label{eq:Qn_H_condition}\ee
Of course, if
\be \wh{Q}_n = -\frac{1}{g(r)} Q_n g(r)
\label{eq:Qhat_false_condition}
\ee
and 
\be \left[Q_n, g(r) \mc{H}_n\right] = 0,
\label{eq:Qn_H_commutator}\ee
then \eqref{eq:Qn_H_condition} is satisfied, but since $Q_n$ is a higher order operator, we have
\be \wh{Q}_n = \wh{\mc{D}}^-_{n} \wh{Q}_{n-k} \wh{\mc{D}}^+_{n-k},
\ee
and there is no guarantee that 
\be \wh{Q}_n = -\frac{1}{g(r)} Q_n g(r)\ee
for any $g(r)$. However, we only require the condition
\be \wh{\mc{D}}^\pm_{n} = -\frac{1}{f(r)} \mc{D}^\mp_{n\pm k} f(r),
\label{eq:ladder_hat_condition}
\ee
where $f(r)$ is the function from \eqref{eq:ladder_vs_action}. Since the \textit{ladder operators} are first order, an expression of this form will always hold, but the requirement on $f(r)$ is key, as it implies
\be \wh{\mc{D}}^\pm_{n} \mc{H}_n = 
-\frac{1}{f(r)} \mc{D}^\mp_{n\pm k} f(r) \mc{H}_n
= -\mc{H}_{n\mp k} \mc{D}^\mp_{n\pm k}.\ee
Thus \eqref{eq:ladder_hat_condition} allows us to use \eqref{eq:abstract_ladder} to push $\mc{H}_n$ to the left past the series of $\wh{\mc{D}}^+$ operators down to $\wh{Q}_0$, where it will have become $\mc{H}_0$, and we can use \eqref{eq:Q0_H_commutator} to move it past $\wh{Q}_0$. Then we can use \eqref{eq:ladder_hat_condition} again to keep moving $\mc{H}_0$ to the right past the $\wh{\mc{D}}^-$ operators, until it becomes $\mc{H}_n$ again, having picked up an overall minus sigh from the step where it was pushed past $\wh{Q}_0$.

To see a specific example, we take $n = 1$ and $k=1$ and compute 
\begin{align}
\wh{Q}_1 \mc{H}_1
&= \wh{\mc{D}}^-_1 \wh{Q}_0 \wh{\mc{D}}^+_0 \mc{H}_1
\nonumber\\
&= -\wh{\mc{D}}^-_1 
\wh{Q}_0 \mc{H}_0\mc{D}^-_1 \nonumber\\
&= \wh{\mc{D}}^-_1
\mc{H}_0 Q_0 \mc{D}^-_1 \nonumber\\
&= -\mc{H}_1 \mc{D}^+_0 Q_0 \mc{D}^-_1 \nonumber\\
&= -\mc{H}_1 Q_1.
\end{align}

Let us now consider the two cases we addressed in this work. For the Reissner--Nordstr\"{o}m black hole, we replace $n$ with $\ell$. The specific dependence of $V_\ell(r)$ on $\ell$ in the form $\ell(\ell+1)$ made the lowering operator for $\ell = 0$ commute with $H_0$, so, using \eqref{eq:Q0H0} and \eqref{eq:ladder}, we have
\be Q_0 = D_0^- = \D \del_r, \quad g(r) = f(r) = \D.\ee
Since $f(r) = g(r)$, in this case we do actually satisfy \eqref{eq:Qhat_false_condition} and \eqref{eq:Qn_H_commutator}. This can also be seen directly from the expression
\be Q_\ell = (-1)^\ell \D^{-(\ell+1)/2} 
\left(\D^{3/2}\del_r\right)^{2\ell+1} \D^{-\ell/2},\ee
where all derivatives act on everything to the right. 

For the conformally coupled scalar in de Sitter, $n$ is replaced by $\g$ and $\ell$, though as we did in the main text, we shall suppress the $\g$ index. As with the black hole, the dependence of $V_{\ell}$ on $\ell$ in the form $\ell(\ell+1)$ gave us a $Q_0$ from the bottom of the ladder. Using \eqref{eq:Q0H0_conf} and \eqref{eq:conf_ladder}, we have
\be Q_0 = \wt{D}_0^- = -\frac{\D}{r^3}\del_r r, 
\quad g(r) = f(r) = \frac{\D}{r^4}.\ee
Once again, the equality of $f(r)$ and $g(r)$ makes this case more tractable, and we have a compact expression for $Q_\ell$:
\be Q_\ell = (-1)^{\ell+1} 
\left(\frac{\D}{r^2}\right)^{-(\ell+1)/2} r^{\ell}
\left(\frac{\D^{3/2}}{r^4}\del_r\right)^{2\ell+1} 
\left(\frac{\D}{r^2}\right)^{-\ell/2} r^{\ell+1}.\ee

\section{Identities for General First Order Differential Operators}
\label{app:identities}

Consider a linear differential operator given by
\be \d_i = A(t,r)x^i + B(t,r) \del_{x^i} + C(t,r) x^i \del_t
+ D(t,r) x^i \del_r.
\ee
If $c$ is an $n$-index traceless symmetric tensor, then
\be c_{i_1 \cdots i_n} \d_{i_1} \cdots \d_{i_n} f(t,r)
= c_{i_1 \cdots i_n} x^{i_1} \cdots x^{i_n} \mc{D}_{n-1} \mc{D}_{n-2} \cdots \mc{D}_1 \mc{D}_0 f(t,r),
\label{eq:delta_id1}
\ee
where
\be \mc{D}_j = \mc{E} + \frac{j}{r} D(t,r),\ee
with
\be \mc{E} = A(t,r) 
+ \left(B(t,r) + \frac{1}{r}D(t,r)\right)\del_r
+ C(t,r) \del_t.\ee
In addition, we find, suppressing the $t$ and $r$ arguments to save space,
\be 
\begin{split}
\label{eq:delta_id2}
&\d^{ij} \d_i \d_j\left(c_{i_1 \cdots i_n} x^{i_1} \cdots x^{i_n} f\right)
= c_{i_1 \cdots i_n} x^{i_1} \cdots x^{i_n} 
\bigg[-\frac{n(n+1)}{r^2}B^2 f 
\\
&\quad
+ r^2 \left(\mc{E} 
+ \frac{n+3}{r^2} B
+ \frac{n+1}{r}D\right) \circ 
\left(\mc{E} 
+ \frac{n}{r^2} B
+ \frac{n}{r}D\right) f
\bigg].
\end{split}
\ee

In the black hole case, with $\d_i = \d_{K_i}$, we have 
\be A(t,r) = \frac{\D'}{2r}, \quad 
B(t,r) = -\frac{r\D'}{2}, \quad
C(t,r) = 0, \quad 
D(t,r) = \frac{1}{2r^2}\del_r\left(r^2 \D\right),\ee
so we find
\be \mc{E} = \frac{\sqrt{\D}}{r}\del_r \sqrt{\D}.\ee
We can express \eqref{eq:delta_id1} as
\be
c_{i_1 \cdots i_n} \d_{i_1} \cdots \d_{i_n} f(t,r)
= c_{i_1 \cdots i_n} x^{i_1} \cdots x^{i_n} 
\frac{1}{r^n \D^{(n+1)/2}}
\left(\D^{3/2} \del_r\right)^n \sqrt{\D} f(t,r),
\label{eq:delta_id1_RN}
\ee
and \eqref{eq:delta_id2} as
\be
\begin{split}
\label{eq:delta_id2_RN}
\d^{ij} \d_i \d_j
\left(c_{i_1 \cdots i_n} x^{i_1} \cdots x^{i_n} f\right)
&= c_{i_1 \cdots i_n} x^{i_1} \cdots x^{i_n} 
\left[
\D^{3/2} \del_r^2\left(\sqrt{\D} f\right)
+ 2 n \D^{3/2} \del_r\left(\frac{\sqrt{\D}}{r} f\right)
\right.\\
&\quad\left.
- \frac{n(n+1)}{4}
\left(\left(\D'\right)^2 - \frac{4\D^2}{r^2}\right) f\right].
\end{split}
\ee

In the de Sitter case with generic coupling, for $\d_i = \d_{K_i}$ or $\d_{P_i}$, we have
\be A(t,r) = 0, \quad 
B(t,r) = L e^{\pm t/L}\frac{\sqrt{\D}}{r^2}, \quad
C(t,r) = \pm e^{-t/L}\frac{r}{\sqrt{\D}}, \quad 
D(t,r) = 0,\ee
with the plus signs for the $\d_{K_i}$ and the minus signs for the $\d_{P_i}$. Then \eqref{eq:delta_id1} and \eqref{eq:delta_id2} greatly simplify:
\begin{align}
\label{eq:delta_id1_dS}
c_{i_1 \cdots i_n} \d_{i_1} \cdots \d_{i_n} f(t,r)
&= c_{i_1 \cdots i_n} x^{i_1} \cdots x^{i_n} \left(L e^{\pm t/L}\frac{\sqrt{\D}}{r^2} \del_r
\pm e^{-t/L}\frac{r}{\sqrt{\D}} \del_t\right)^n f(t,r)\\
\begin{split}
\label{eq:delta_id2_dS}
\d^{ij} \d_{i}\d_{j} 
\left(c_{i_1 \cdots i_n} x^{i_1} \cdots x^{i_n} f\right)
&= c_{i_1 \cdots i_n} x^{i_1} \cdots x^{i_n}
\left[r^2 \left(\pm e^{\pm t/L}\frac{r}{\sqrt{\D}} \del_t + L e^{\pm t/L}\frac{\sqrt{\D}}{r^2} \del_r\right)^2 f
\right. \\
&\quad \left.
+ \left(2n + 3\right) L e^{\pm t/L} \frac{\sqrt{\D}}{r} \left(\pm e^{-t/L}\frac{r}{\sqrt{\D}} \del_t 
+ L e^{\pm t/L}\frac{\sqrt{\D}}{r^2} \del_r\right)f\right].
\end{split}
\end{align}

Lastly, in the de Sitter case with conformal coupling, for $\d_i = \d_{\wt{K}_i}$, we have
\be A(t,r) = -\frac{1}{L}, \quad B(t,r) = L, \quad
C(t,r) = 0, \quad 
D(t,r) = -\frac{r}{L},\ee
and we find
\begin{align} 
c_{i_1 \cdots i_n} \d_{i_1} \cdots \d_{i_n} f(t,r)
&= c_{i_1 \cdots i_n} x^{i_1} \cdots x^{i_n} 
\frac{L^n r^{n+1}}{\D^{(n+1)/2}}
\left(\frac{\D^{3/2}}{r^4} \del_r\right)^n 
\frac{\sqrt{\D}}{r} f(t,r).
\label{eq:delta_id1_dS_conf}\\
\begin{split}
\label{eq:delta_id2_dS_conf}
\d^{ij} \d_{i}\d_{j} 
\left(c_{i_1 \cdots i_n} x^{i_1} \cdots x^{i_n} f\right)
&= 
c_{i_1 \cdots i_n} x^{i_1} \cdots x^{i_n}
\left[\frac{L^2}{r^4} \D^{3/2} \bigg(\del_r^2\left(\sqrt{\D} f\right)
\right.\\
&\quad
\left.
+ \frac{2n}{r^{(n+1)/2}} 
\del_r\left(r^{(n-1)/2}\sqrt{\D} f\right)
\bigg)
- n(n+1)\frac{L^2}{r^2} f
\right].
\end{split}
\end{align}

\section{More Ways to Climb Ladders}
\label{app:ladder_extensions}

Though the focus of this work has been on operators that act on the radial $\f(r)$ solutions, we have seen that we can act with various (C)KVs to move around the space of the full $\F(t,r,\th,\varphi)$ solutions. In this appendix, we shall explore this further.

We first need one more tool. For both the black hole and de Sitter space, we have a triplet of KVs $J_i$ corresponding to spatial rotations. In the usual way, we can construct operators that raise and lower $m$. We take
\be J_{\pm}^\m \del_\m 
= \left(J_1^\m \pm i J_2^\m\right)\del_\m 
= \pm i e^{\pm i\f} \del_\th - e^{\pm i\f} \cot\th\, \del_\varphi.\ee
Then in the de Sitter case, we find
\be\d_{J_{\pm}} \F_{\g\ell m}(t,r,\th,\varphi) 
\propto \F_{\g\ell, m\pm 1}(t,r,\th,\varphi).\ee
(For the black hole case, simply drop the $\g$ index and the $t$ argument.) 

The power of these new operators comes from the fact that $\F_{\g\ell m} = 0$ for $\ell < m$. Focusing on the black hole for now, we see that \eqref{eq:RN_delta_K3} shows that for $m = \ell$, the effect of $\d_{K_3}$ on $\F_{\ell\ell}$ is solely to raise $\ell$, as the second term vanishes. We can thus map any $\F_{\ell m}$ to $\F_{\ell+1,m}$, up to an overall constant, with the following sequence of operators
\be \left(\d_{J_-}\right)^{\ell-m} \d_{K_3} \left(\d_{J_+}\right)^{\ell-m} \F_{\ell m} \propto \F_{\ell+1, m}, \ee
or perhaps more clearly,
\be
(\ell,m)
\xrightarrow{\left(\d_{J_+}\right)^{\ell-m}}
(\ell+1,\ell)
\xrightarrow{\d_{K_3}}
(\ell+1,\ell)
\xrightarrow{\left(\d_{J_-}\right)^{\ell-m}}
(\ell+1,m).
\ee
Unfortunately, in the black hole case there is no way to map $\F_{\ell m}$ to $\F_{\ell-1,m}$ using any combination of (C)KVs.

Moving to de Sitter space, we find from \eqref{eq:dS_delta_K3} that setting $\g = \g_\pm$ makes the $F^+_{\g\ell}$ term vanish, so the effect of $\d_{K_3}$ is to lower $\ell$ alone. This gives us a way to map any quasinormal mode $\F_{\g\ell m}$ to $\F_{\g,\ell-1,m}$. We lower $\g = \g_{\pm} + 2n$ to $\g_{\pm}$ with $n$ applications of $\d^{ij} \d_{K_i} \d_{K_j}$, lower $\ell$ with $\d_{K_3}$, and then raise $\g$ back up with $n$ applications of $\d^{ij} \d_{P_i} \d_{P_j}$:
\be \left(\d^{ij} \d_{P_i} \d_{P_j}\right)^n \d_{K_3} \left(\d^{kl} \d_{K_k} \d_{K_l}\right)^n \F_{\g_{\pm}+2n,\ell m} 
\propto \F_{\g_{\pm}+2n,\ell-1,m},\ee
or,
\be
(\g_{\pm}+2n,\ell,m) 
\xrightarrow{\left(\d^{kl} \d_{K_k} \d_{K_l}\right)^n}
(\g_{\pm},\ell,m)
\xrightarrow{\d_{K_3}}
(\g_{\pm},\ell-1,m)
\xrightarrow{\left(\d^{ij} \d_{P_i} \d_{P_j}\right)^n}
(\g_{\pm}+2n,\ell-1,m).
\ee
Raising $\ell$ is more involved but can be done, unlike in the Reissner–Nordstr\"{o}m case. We must first lower $\g$ to $\g_\pm$ and $m$ to zero with $\d^{ij} \d_{K_i} \d_{K_j}$ and $\d_{J_-}$ (in either order). We can then lower $\ell$ to zero with $\d_{K_3}$. Then we use 
\be c^{\ell+1, m}_{i_1\cdots i_{\ell+1}} 
\d_{P_{i_1}} \cdots \d_{P_{i_{\ell+1}}}\ee
to jump from $(\g_\pm,0,0)$ all the way to $(\g_\pm,\ell,m)$. Lastly, we use $\d^{ij} \d_{P_i} \d_{P_j}$ to raise $\g$ back to its original value. We can represent this as
\be
\left(\d^{ij} \d_{P_i} \d_{P_j}\right)^n
\left(c^{\ell+1, m}_{i_1\cdots i_{\ell+1}} 
\d_{P_{i_1}} \cdots \d_{P_{i_{\ell+1}}}\right)
\left(\d_{K_3}\right)^\ell
\left(\d^{ij} \d_{K_i} \d_{K_j}\right)^n
\left(\d_{J_-}\right)^m \F_{\g_{\pm}+2n,\ell m}
\propto \F_{\g_{\pm}+2n,\ell+1,m},
\ee
or
\be
\begin{split}
&(\g_{\pm}+2n,\ell,m) 
\xrightarrow{\left(\d_{J_-}\right)^m}
(\g_{\pm}+2n,\ell,0)
\xrightarrow{\left(\d^{ij} \d_{K_i} \d_{K_j}\right)^n}
(\g_{\pm},\ell,0)
\xrightarrow{\left(\d_{K_3}\right)^\ell}\\
&(\g_{\pm},0,0)
\xrightarrow{
c^{\ell+1, m}_{i_1\cdots i_{\ell+1}} 
\d_{P_{i_1}} \cdots \d_{P_{i_{\ell+1}}}
} 
(\g_{\pm},\ell+1,m)
\xrightarrow{\left(\d^{ij} \d_{P_i} \d_{P_j}\right)^n}
(\g_{\pm}+2n,\ell+1,m).
\end{split}
\ee
Thus in the de Sitter case, we can move from any quasinormal mode to any other via a series of operations, all built on KVs.

\bibliographystyle{utphys}
\bibliography{refs}
\end{document}